\documentclass[a4paper,twocolumn,11pt,accepted=2025-04-23]{quantumarticle}
\pdfoutput=1
\usepackage[utf8]{inputenc}
\usepackage[english]{babel}
\usepackage[T1]{fontenc}
\usepackage{amsmath}
\usepackage{hyperref}

\usepackage{tikz}
\usepackage{lipsum}

\begin{document}

\title{Building a fusion-based quantum computer using teleported gates} 

\author{Ashot Avanesov}
\affiliation{
Russian Quantum Center, Russia, Moscow, 121205, Bol'shoy bul'var 30 building 1
}
\email{avanesov@phystech.edu}
\author{Alexander Shurinov}
\affiliation{
 Quantum Technologies Centre, Lomonosov Moscow State University, Russia, Moscow, 119991, Leninskie Gory 1 building 35
}
\author{Ivan Dyakonov}
\affiliation{
 Quantum Technologies Centre, Lomonosov Moscow State University, Russia, Moscow, 119991, Leninskie Gory 1 building 35
}
\affiliation{
Russian Quantum Center, Russia, Moscow, 121205, Bol'shoy bul'var 30 building 1
}
 \email{dyakonov@quantum.msu.ru}
\author{Stanislav Straupe}
\affiliation{
 Quantum Technologies Centre, Lomonosov Moscow State University, Russia, Moscow, 119991, Leninskie Gory 1 building 35
}
\affiliation{
Russian Quantum Center, Russia, Moscow, 121205, Bol'shoy bul'var 30 building 1
}
\maketitle

\begin{abstract}
We adopt a method of the quantum gate teleportation for converting
circuit-based quantum computation primitives into fusion networks.
By using the presented scheme for the CNOT gate we construct translation of the circuit for the foliated surface code into a fault tolerant fusion network. Finally, we construct two new fusion based quantum computation models and study their fault tolerance properties.
\end{abstract}

\newcommand{\ket}[1]{|#1\rangle}
\newcommand{\bra}[1]{\langle#1|}

\section{Introduction}

The quantum algorithms are typically considered as a unitary transformation of a qubit register followed by measurement of each qubit. The unitary transformation is typically represented as a sequence of single- and two-qubit gates and it is usually the most convenient method to design a new algorithm. The single- and two-qubit gates are drawn from a finite set of gates called a universal set that has the following property: an arbitrary unitary operation over a fixed number of qubits can be approximated by a composition of quantum gates from a universal set \cite{kitaev_solovey,elementary_gates}. We will call this approach a Circuit Based Quantum Computation (CBQC) model \cite{nielsen_chang}. The key steps in CBQC include initialization of a register, applying a sequence of quantum gates which defines a particular algorithm, and quantum measurements that produce outcomes of the computation.  

Besides CBQC there exist other models of quantum computation \cite{annealing,adiabatic_computation,anyons,topological_computation,raussendorf_one-way,raussendorf_MBQC,FBQC_1}, and some of them have proven to be more natural for implementation on particular physical platforms. For example, Measurement-Based Quantum Computation (MBQC) \cite{raussendorf_one-way,raussendorf_MBQC} and Fusion-Based Quantum Computation (FBQC) models \cite{FBQC_1,FBQC_interleaving,FBQC_logical_blocks,FBQC_unifying_with_ZX,FBQC_increasing_error_tolerance} are particularly well suited for photonic architectures \cite{fusion_types,qudits_computation,percolation_mbqc,gimeno-segovia_3ghz,omkar1,omkar2,zhang_compilation} because the coherent evolution of a quantum register is substituted with measurement and teleportation. The aforementioned models are equivalent to  the CBQC and are able to perform arbitrary quantum algorithm. We refer to \cite{raussendorf_one-way} for the proof of CBQC and MBQC equivalence.

In MBQC an algorithm instance is executed in two steps. First, a large multi-qubit entangled state $\ket{\Phi_{\mathcal{C}}}$ is created. The size of the $\ket{\Phi_{\mathcal{C}}}$ depends on corresponding CBQC circuit complexity. Afterwards, the entire state undergoes a sequence of adaptive single-qubit measurements. A special type of entangled states -- cluster states \cite{cluster_states} -- serve as a resource for the MBQC-based quantum computing. The MBQC model is seamlessly compatible with modern error correction codes and the outcomes of the single-qubit measurements can also be used to detect and fix errors occurring during the algorithm execution  \cite{raussendorf_fault_tolerant_MBQC,brown_foliation,stace_foliation,beyond_foliation}.

Preparation of the initial resource state $\ket{\Phi_{\mathcal{C}}}$ is the most demanding stage of the MBQC. The number of qubits in $\ket{\Phi_{\mathcal{C}}}$ depends on the algorithm depth, and it is usually impractical to create the whole state at once. The mechanics of the MBQC architecture grant a possibility of conveyor-like algorithm execution. The resource state $\ket{\Phi_{\mathcal{C}}}$ never exists as a whole and only the required fraction $\ket{\phi_{\mathcal{C}}}$ of this state is instantaneously present within a processor. Each step of an algorithm involves measurement of some of the qubits of $\ket{\phi_{\mathcal{C}}}$ and linking of new qubits to the intact qubits from $\ket{\phi_{\mathcal{C}}}$. This strategy is particularly suitable to the linear-optical architecture which suffers from a high-rate of qubit loss. The benefit stems from the fact that a qubit within a processor has a short lifetime and hence there is no need to store photons as long as the algorithm runs.

Photon loss \cite{loss_simulation,loss_simulation_nonuniform,loss_effect} is not the only detrimental factor in a linear-optical system. Entangling operations in dual-rail encoding are intrinsically probabilistic \cite{LOCC,probabilistic_operation}.  The conveyor-like algorithm execution strategy must be aided with an error-correction procedure to successfully overcome both photon loss and limited success probability of entangling operations. The structure of the resource state $\ket{\phi_{\mathcal{C}}}$ defines the fault-tolerance properties to the qubit loss and erroneous entangling operations \cite{stace_loss_correction,stace_high_threshold}.

The initial cluster state of an MBQC-based processor can be assembled using multiple copies of identical few-qubit entangled states, known as resource states. The assembly may be carried out by applying fusion gates to entangle separate resource states. In this work we focus on a Type-II fusion gate \cite{fusion_types} (also called \textit{fusion} as a shorthand notation), which is a projective measurement in the two-qubit Bell basis. Introduction of fusion of resource states in order to build up large-scale quantum correlations is a cornerstone of the FBQC model. The FBQC model goes beyond the MBQC and uses fusion gates also to induce the quantum state transfer between the qubits and apply unitary operations prescribed by an algorithm of choice. Then the algorithm implementation in the FBQC model is divided into two parts. Firstly, the resource states are generated and afterwards the reconfigurable fusion gates are applied to the resource states. These two steps may be repeated continuously until the measurement of the algorithm's outcome is performed. A specific FBQC model is determined by the fusion network, which represents the mutual configuration of resource states and fusions. A graph with two distinct sets of edges conveniently represents a fusion network: the first set contains graphs corresponding to the resource states, and the second one indicates the locations of the fusion gates between the resource qubits. Some fusion networks possess intrinsic fault-tolerance features (fault-tolerant fusion networks)\cite{FBQC_1} which allows to use fusion gate outcomes for error detection and application of an error-correction protocol.

The work \cite{FBQC_1} reports two types of fusion networks based on different resource states: a 4-qubit star and a 6-qubit hexagon cluster states. The first network can be considered as a simple modification of the Raussendorf model, which is the MBQC architecture based on Fault-Tolerant Cluster State produced from the foliation \cite{stace_foliation,brown_foliation} of the surface code \cite{kitaev_toric,kitaev_codes,planar_code,topological_quantum_memory,fowler_surface_codes}. The similar approach was used in the paper \cite{floquet_QC}, where the FBQC scheme was constructed from the foliation of the Floquet Color Code \cite{floquet_codes,color_code,floquet_measurement}. The construction of the second model from \cite{FBQC_1} was explained in \cite{FBQC_unifying_with_ZX}, where the authors introduced a general framework for the description of fusion networks based on ZX-calculus \cite{zx_calculus_1,zx_calculus_2,zx_calculus_review}.

Here we consider a direct path to construct a fusion network from quantum circuit using the quantum gate teleportation principle \cite{teleportation,quantum_gate_teleportation,qgt_universal,qgt_computation,qgt_for_qc,qgt_fault-tolerant,qgt_derivation_for_mbqc,qgt_mbqc_intro}. In \cite{quantum_gate_teleportation}, a scheme was proposed, where the quantum algorithm is decomposed into a sequence of elementary operations which are implemented using the quantum gate teleportation protocol. One can introduce a fusion network that describes the resulting configuration of operations, where the Bell measurements and the Bells states correspond to the fusions and the resource states, respectively. Thus, we have the opportunity to establish a connection between the decomposition into a sequence of gate teleportation protocols of elementary gates and a fusion network, providing a direct translation between the CBQC and the FBQC models. In Sec.~\ref{sec: quantum gate teleportation}, we revise the approach of \cite{quantum_gate_teleportation} and specify the equivalent fusion circuits for the Hadamard, T- and CNOT gates. 
In Sec.~\ref{sec: fusion networks}, this translation technique between CBQC and FBQC operations is demonstrated using the examples of a foliated surface code. By applying this method, we construct previously unreported fusion networks. In Sec.~\ref{sec: noise tolerance}, we investigate their fault-tolerant properties.

\section{Quantum gate teleportation}
\label{sec: quantum gate teleportation}

Fusion is a destructive measurement of commutative Pauli observables. Here we consider a variant of a Type-II fusion gate \cite{fusion_types}, which is a two-qubit projective measurement determined by the group $\langle X_{q_1}Z_{q_2},\ Z_{q_1}X_{q_2}\rangle$, where $q_1$ and $q_2$ are qubit indices. The operators $X_i$ and $Z_i$ are $X$ and $Z$ Pauli matrices acting in the Hilbert space of an $i$-th qubit. The state of the measured qubits is projected onto the following basis states:
\begin{equation}
\label{eq: XZ ZX basis states}
    \begin{array}{c}
    \ket{\underline{0,\ 0}}_{q_1q_2}=\frac{1}{\sqrt{2}}(\ket{+}_{q_1}\ket{0}_{q_2}+\ket{-}_{q_1}\ket{1}_{q_2}), \\
    \ket{\underline{0,\ 1}}_{q_1q_2}=\frac{1}{\sqrt{2}}(\ket{+}_{q_1}\ket{0}_{q_2}-\ket{-}_{q_1}\ket{1}_{q_2}), \\
    \ket{\underline{1,\ 0}}_{q_1q_2}=\frac{1}{\sqrt{2}}(\ket{+}_{q_1}\ket{1}_{q_2}+\ket{-}_{q_1}\ket{0}_{q_2}), \\
    \ket{\underline{1,\ 1}}_{q_1q_2}=\frac{1}{\sqrt{2}}(\ket{+}_{q_1}\ket{1}_{q_2}-\ket{-}_{q_1}\ket{0}_{q_2}),
    \end{array}
\end{equation}
where $\ket{\pm}=\frac{1}{\sqrt{2}}(\ket{0}\pm\ket{1})$ and

\begin{equation}
\begin{array}{c}
    X_{q_1}Z_{q_2}\ket{\underline{u,\ v}}_{q_1q_2}=(-1)^u\ket{\underline{u,\ v}}_{q_1q_2},\\
    Z_{q_1}X_{q_2}\ket{\underline{u,\ v}}_{q_1q_2}=(-1)^v\ket{\underline{u,\ v}}_{q_1q_2}.
\end{array}
\end{equation}

One can also notice that

\begin{equation}
\label{eq: XZ ZX basis states in simple form}
    \ket{\underline{u,\ v}}_{q_1q_2} = X^v_{q_1}X^u_{q_2}\ket{\underline{0,\ 0}}_{q_1q_2}.
\end{equation}

The key elements of a generic FBQC processor are the resource state generator (RSG) and fusion modules, connecting the neighbouring RSG-generated states. The fusion gate decouples the measured qubits from the logical register or completely destroys them in case of photonic-encoded qubits. Then the RSG may either reuse the decoupled qubits or generate the new ones. Hence the build-up of a fusion network can be also performed in time dimension which requires generation of new resource states and applying fusions between resource state generated at different moments in time. Therefore the computation process in the FBQC model naturally splits into a sequence of cycles. Each cycle the RSG network produces the resource states. Some of their qubits are fused with each other and with the qubits generated during several previous cycles. A set of remaining qubits (i.e. the qubits that are not fused in the current cycle) is stored until fusion with resource states from future cycles is performed. This process can be viewed as quantum gate teleportation \cite{quantum_gate_teleportation}: a required unitary transformation is applied to a newly generated qubit via a teleportation protocol \cite{teleportation}.

We consider the translation of a CBQC quantum circuit to a fusion network, that follows a simple strategy, which actually was described in \cite{quantum_gate_teleportation}. First, the quantum circuit is decomposed into a sequence $\mathcal{S}$ of elementary operations $g_{i}$ drawn from a universal gate set. Then, a quantum gate teleportation protocol $\mathcal{T}_{g_{i}}$ for each elementary operation is defined. The gate teleportation protocol includes a resource state, a Bell-type measurement and single-qubit gates. Substituting each $\mathcal{T}_{g_{i}}$ into the initial gate sequence $\mathcal{S}$ completes the translation. The resulting configuration is the fusion network that implements the computation.

\begin{figure}
    \centering
    \includegraphics[scale=0.6, width=\linewidth]{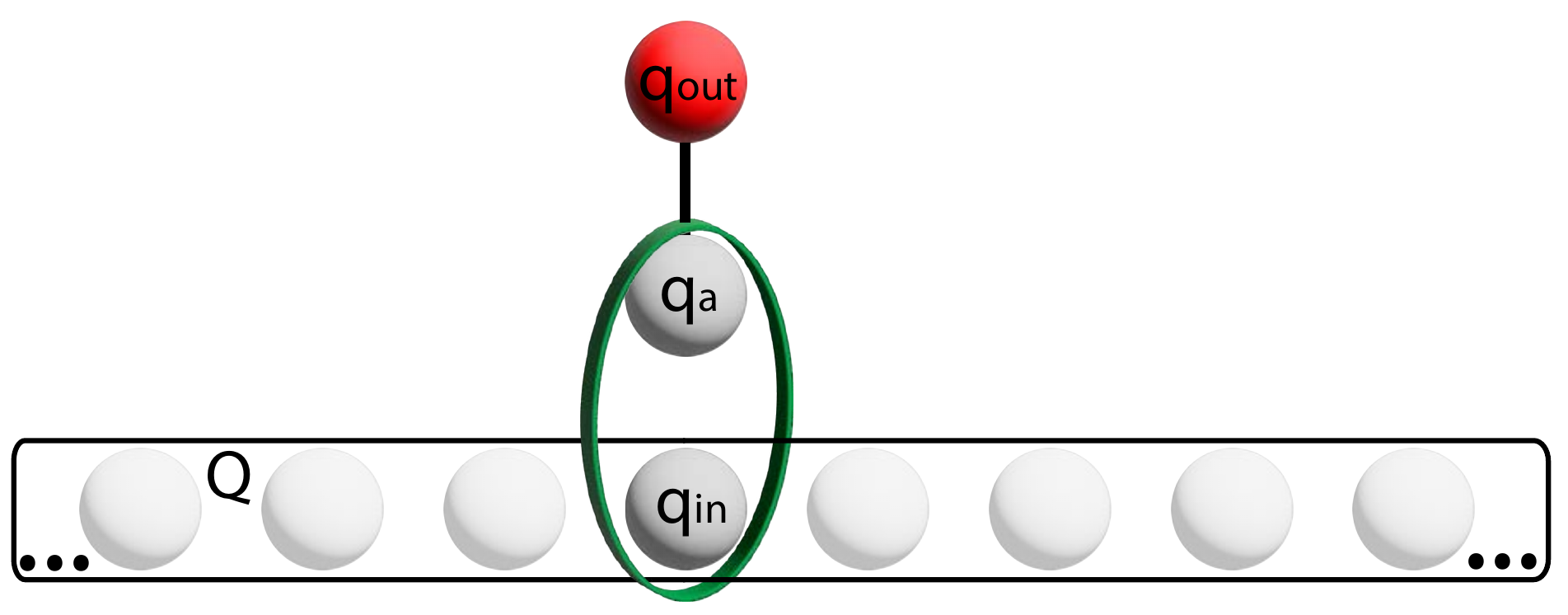}
    \caption{Quantum gate teleportation scheme for implementation of a single-qubit unitary operation. Qubit $q_{in}$ from the initial array of qubits $Q$ is fused with qubit $q_a$ of the ancillary state.}
    \label{fig:1qgate}
\end{figure}

Let us now adapt the gate teleportation \cite{quantum_gate_teleportation} protocol to implement an arbitrary quantum circuit using $\langle X_{q_1}Z_{q_2},\ Z_{q_1}X_{q_2}\rangle$ fusion gates. We start with the single-qubit gates. The procedure is illustrated in Fig.~\ref{fig:1qgate}. In the input, we have an array $Q$ of qubits which are in some state $|\psi\rangle$. In order to perform a unitary operation on a single qubit from $Q$ one uses a two-qubit ancilla state. Let us denote the fusing qubit from the initial system $Q$ as $q_{in}$ and a pair of ancillary qubits as $q_a$ and $q_{out}$. The state of qubits $q_a$ and $q_{out}$ is

\begin{equation}
\label{eq: ancilla for single qubit gate}
|anc\rangle = U_{q_{out}} CZ_{q_{out},q_a}|+\rangle_{q_{out}}|+\rangle_{q_a}.
\end{equation}
Here, $U_{q_{out}}$ is a target unitary operation that is performed on an ancillary qubit $q_{out}$. It is worth noting that $CZ_{q_{out},q_a}|+\rangle_{q_{out}}|+\rangle_{q_a}$ is a Bell cluster state. After fusing qubits $q_{in}$ and $q_a$, only the qubit $q_{out}$ remains in the register and has the state $U_{q_{out}}|\psi\rangle.$ Indeed, the state of the register before the fusion is

\begin{equation}
    \ket{anc}(\ket{0}_{q_{in}}\ket{\psi_0}_{Q\setminus\{q_{in}\}}+\ket{1}_{q_{in}}\ket{\psi_1}_{Q\setminus\{q_{in}\}}).
\end{equation}
Here $\ket{\psi_0}$ and $\ket{\psi_1}$ are unnormalized vectors from a Hilbert space of the set $\tilde{Q} =Q\setminus\{q_{in}\}$ such that $\ket{\psi} = \ket{0}_{q_{in}}\ket{\psi_0}_{\tilde{Q}}+\ket{1}_{q_{in}}\ket{\psi_1}_{\tilde{Q}}$. Substituting the expression for $\ket{anc}$ from (\ref{eq: ancilla for single qubit gate}) we get

\begin{equation}
\label{eq: state for single qubit gate}
    \ket{anc}\ket{\psi}=\frac{U_{q_{out}}H_{q_a}}{\sqrt2}\sum_{k,\ l=0}^1\ket{k}_{q_{out}}\ket{\psi_l}_{\tilde Q} \ket{k}_{q_a}\ket{l}_{q_{in}},
\end{equation}
where $H_{q_a}$ is the Hadamard operation that acts on the qubit $q_a$. Then, we project the state (\ref{eq: state for single qubit gate}) onto the basis (\ref{eq: XZ ZX basis states}) $\left\{\ket{\underline{u,\ v}}_{q_a,\ q_{in}}:\ u,\ v\ \in \{0,\ 1\}\right\}$, and obtain the following result (the $(u,\ v)$ outcomes correspond to a correction matrix) %

\begin{equation}
    \begin{array}{c}
    \bra{\underline{u,\ v}}_{q_aq_{in}}\ket{anc}\ket{\psi}\sim\\ \\ \sim U_{q_{out}}Z^v_{q_{out}}X^u_{q_{out}}\ket{\psi}_{\tilde Q\cup \{q_{out}\}},
    \end{array}
\end{equation}
where $\ket{\psi}_{\tilde Q\cup \{q_{out}\}} = \ket{0}_{q_{out}}\ket{\psi_0}_{\tilde Q} + \ket{1}_{q_{out}}\ket{\psi_1}_{\tilde Q}$.

\begin{figure}
    \centering
    \includegraphics[scale=0.6, width=\linewidth]{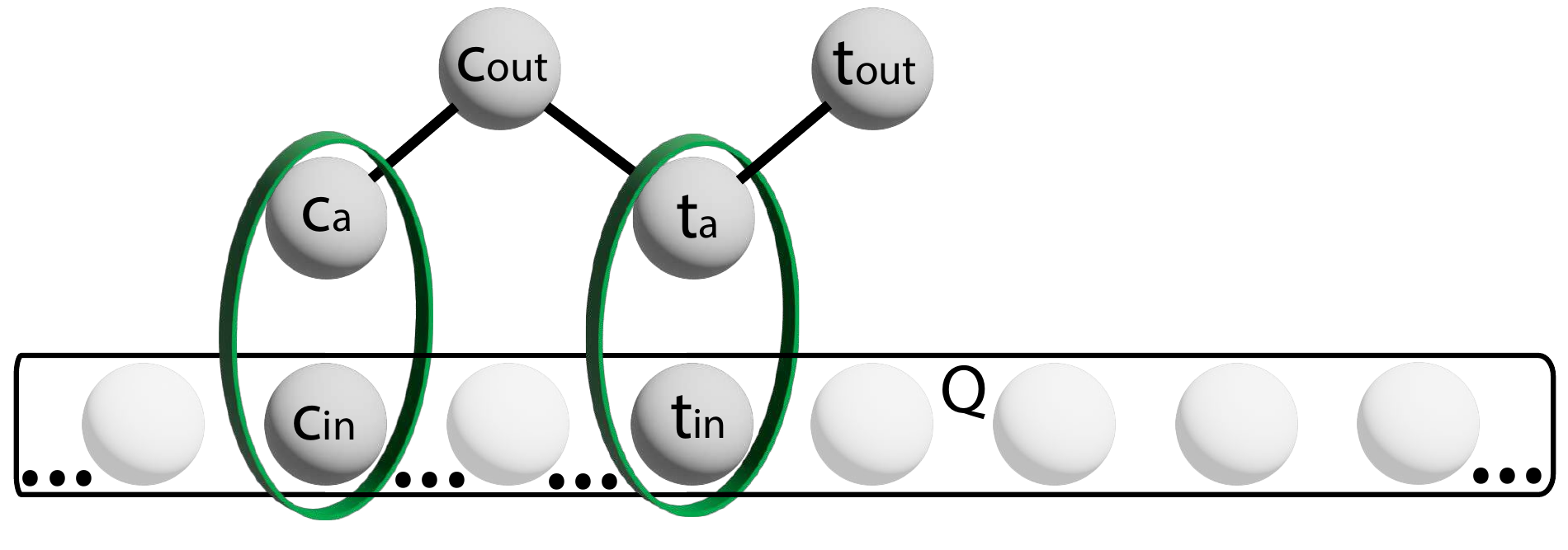}
    \caption{Quantum gate teleportation scheme for implementation of $CNOT$ operation. Qubits $c_{in}$ and $t_{in}$ from initial array of qubits $Q$ are fused with qubit $c_a$ and $t_a$ of the $4$-qubit ancilla state.}
    \label{fig:2qgate}
\end{figure}
The byproduct matrices $Z_{q_{out}}$ and $X_{q_{out}}$ can be commuted through the subsequent operations to the end of the quantum circuit. There, they can be accounted for by modifying the basis of the final measurements.

A two-qubit gate implementation follows the same recipe and requires at least two fusion gates and a four-qubit ancillary state. We exemplify the idea by showing the exact procedure for the CNOT gate teleportation. The initial state of the register is

\begin{equation}
    \ket{\psi} = \sum_{k,\ l}\ket{k}_{c_{in}}\ket{l}_{t_{in}}\ket{\psi_{kl}}_{\tilde Q}.
\end{equation}
Here, $\tilde Q = Q\setminus\{c_{in},\ t_{in}\}$ denotes the set of qubits in $Q$ excluding the control qubit $c_{in}$ and the target qubit $t_{in}$. Though, we do not need to require that the qubits $t_{in}$ and $c_{in}$ were within some neighborhood, in practical fault-tolerant architectures the locality can be important feature. We label the ancillary state qubits $c_a$, $t_a$, $c_{out}$, $t_{out}$. The qubits $t_{in}$ and $c_{in}$ are fused with the qubits $c_a$ and $t_a$.  We express the ancillary state in the following way

\begin{equation}
\label{eq: cnot ancilla state}
    \ket{anc} = \sum_{k,\ l} H_{c_a}\ket{k}_{c_a}H_{t_a}\ket{l}_{t_a}\ket{\alpha_{kl}}_{c_{out},\ t_{out}},
\end{equation}
where $H_{c_a}$ and $H_{t_a}$ are the Hadamard gates and $\ket{\alpha_{kl}}_{c_{out},\ t_{out}}$ is a two qubit state. We fuse two pairs of qubits: $(c_{in},\ c_a)$ and $(t_{in},\ t_a)$. According to (\ref{eq: XZ ZX basis states}), the corresponding operation projects $\ket{anc}\ket{\psi}$ onto the states

\begin{equation}
\begin{array}{c}
\ket{\underline{u_c,\ v_c}}_{c_{a},\ c_{in}}\ket{\underline{u_t,\ v_t}}_{t_{a},\ t_{in}}=\\ \\=X^{v_c}_{c_a}X^{u_c}_{c_{in}}X^{v_t}_{t_a}X^{u_t}_{t_{in}}\ket{\underline{0,\ 0}}_{c_a,\ c_{in}}\ket{\underline{0,\ 0}}_{t_a,\ t_{in}}.
\end{array}
\end{equation}

After the fusion measurements, we obtain

\begin{widetext}
\begin{equation}
\begin{array}{c}
    \bra{\underline{u_c,\ v_c}}_{c_{a},\ c_{in}}\bra{\underline{u_t,\ v_t}}_{t_{a},\ t_{in}}\ket{anc}\ket{\psi}\sim\\ \\
    \sim \sum_{k,\ l}(-1)^{(k+u_c)v_c}(-1)^{(l+u_t)v_t}\ket{\alpha_{k+u_c,\ l+u_t}}_{c_{out},\ t_{out}}\ket{\psi_{kl}}_{\tilde Q}.
\end{array}
\end{equation}
\end{widetext}

We demand that in the case of $u_c=u_t=v_c=v_t=0$ outcome the resulting state must undergo the CNOT operation without any correction needed. Using this condition we get the following expression for $\ket{\alpha_{kl}}_{c_{out},\ t_{out}}$:
\begin{equation}
    \ket{\alpha_{kl}}_{c_{out},\ t_{out}} = CNOT_{c_{out},\ t_{out}}\ket{k}_{c_{out}}\ket{l}_{t_{out}},
\end{equation}
and finalize the form of the required ancillary state. In the case of an arbitrary outcome, the resulted transformation reads as

\begin{widetext}
\begin{equation}
    \ket{\psi}\to CNOT_{c_{out},\ t_{out}}Z_{c_{out}}^{v_c}X^{u_c}_{c_{out}}Z_{t_{out}}^{v_t}X^{u_t}_{t_{out}}\ket{\psi}_{\tilde Q\cup\{c_{out},\ t_{out}\}}.
\end{equation}
\end{widetext}

Direct check shows that the group $\langle X_{c_{a}}Z_{c_{out}},\ X_{c_{out}}Z_{c_a}Z_{t_a},\ X_{t_a}Z_{c_{out}}Z_{t_{out}},\ X_{t_{out}}Z_{t_a}\rangle$ stabilizes the ancillary state (\ref{eq: cnot ancilla state}) and corresponds to a stabilizer group of a linear four-qubit cluster state, illustrated in Fig.~\ref{fig:2qgate}.

\section{Fusion networks for foliated surface code based on 4-qubit ancillary states.}
\label{sec: fusion networks}

In the previous section, we presented the implementation of arbitrary single-qubit gates and the two-qubit CNOT gate based on the quantum gate teleportation framework. Let us recall that the CNOT and single-qubit gates constitute a universal gate set \cite{kitaev_solovey}. Thus, arbitrary quantum circuit composed of such operations can be efficiently mapped onto fusion network according to the scheme described in the previous section. It is particularly interesting to consider the construction of fusion networks from quantum circuits implementing fault tolerant operations.

The resource states which comprise a fusion network belong to a commutative group $\mathcal{R}$. Fusion measurements are also determined by a commutative group, which we denote as $\mathcal{F}=\langle f_1,...,\ f_m,\ -I\rangle$. Here, $f_1,...,\ f_m$ are the measured Pauli observables, and the measurement outcomes can be either $+1$ or $-1$. As in \cite{FBQC_1}, we include the element $-I$ in the group $\mathcal{F}$. If $\mathcal{F} \cap \mathcal{R} \neq \emptyset$ then certain combinations of observable outcomes must be fixed. It means that if $\pm\prod_{j=1}^m f_j^{k_j}\in\mathcal{R}$, then
\begin{equation}
\label{eq: parity condition}
    \pm\prod_{j=1}^mf_j^{k_j} = 1,
\end{equation}
where we consider $f_j$ as a measurement outcome of a corresponding observable. The fusion network which satisfies $\mathcal{C}=\mathcal{R}\cap\mathcal{F}\neq\emptyset$ is referred to as a fault-tolerant fusion network (FTFN). The condition~(\ref{eq: parity condition}) is essential for further implementation of error correction algorithms. The group $\mathcal{C}$ determines a stabilizer code, and elements of $\mathcal{C}$ are check operators of the code. Each element of the group $\mathcal{C}$ can be expressed as a composition of the generators of the fusion group $\mathcal{F}$. If an error is described by a Pauli operator that anticommutes with a check operator, which is a product of observables $f_j$, it can be detected because the outcomes of the observables $f_j$ in fusion measurements will violate the condition (\ref{eq: parity condition}).

We are interested in the case, where one can choose the generators of $\mathcal{C}$ such that each pair of generators in $\mathcal{C}$ has at most one common element $f_i$ within their decompositions. Then, the group $\mathcal{C}$ can be represented by a syndrome graph. In this graph, the vertices represent generators of $\mathcal{C}$, and the edges correspond to the measured observable outcomes. For any vertex $v$ in the syndrome graph, the generator $c_v$ of the group $\mathcal{C}$ can be expressed as a composition of fusion operators associated with the edges connected to $v$.  The syndrome graph provides a visual representation of the relation between generators in $\mathcal{C}$ and the measured operators, aiding in the construction of error correction procedures.

The transpilation of a fault-tolerant quantum circuit using the proposed fusion-based equivalent of single- and two-qubit gates ends up with an instance of FTFN. In this section, we illustrate the construction of the FTFN specifically for the foliated surface code \cite{topological_quantum_memory,stace_foliation}.%

\begin{figure}
    \centering
    \includegraphics[scale=1.0]{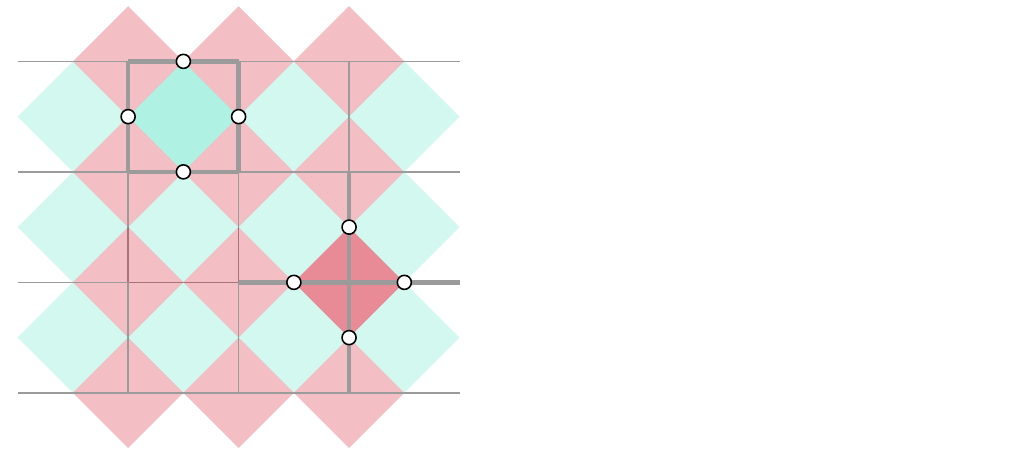}
    \caption{The square lattice of the surface code stabilizers. X-checks and Z-checks are associated with vertices and plaquettes of the lattice . The edges correspond to physical qubits as for highlighted stabilizers. Here, we also introduce squares of two colors for each type of stabilizer check: red for X-checks and blue for Z-checks.} 
    \label{fig:surface code}
\end{figure}

We start with a brief review of the surface code. The surface code is a stabilizer code, its stabilizers are associated with plaquettes and vertices of a square lattice $G$, see Fig.~\ref{fig:surface code}. Every edge of the graph $G$ corresponds to a physical qubit. The stabilizer corresponding to the vertex $v$ of the lattice $G$ is called a star stabilizer and has the form

\begin{equation}
\label{eq: star stabilizer}
    S_v = \prod_{q\in E_{G}(v)}X_q,
\end{equation}
where $E_G(v)$ is a set of edges that are incident to the vertex $v$. It is worth noting that every plaquette in the primal lattice $G$ corresponds to a vertex in the dual lattice $G^*$. The stabilizer associated with the plaquette $v$ from $G^*$ is

\begin{equation}
\label{eq: plaquette stabilizer}
    P_v = \prod_{q\in E_{G^*}(v)}Z_q.
\end{equation}

The surface code is defined as a Hilbert subspace stabilized by both plaquette $P_{v}$ ($Z$-check operator) and star $S_{v}$ ($X$-check operator) stabilizers. In the absence of noise, the measurement of each stabilizer should yield an outcome of $+1$, indicating that the physical qubits are in a state consistent with the parity conditions. However, errors can lead to a violation of the parity condition, causing some of the check operators to output $-1$ values after the measurement. The set of these faulty outcomes forms the error syndrome, which serves as an input to a decoding algorithm that reconstructs the error and prescribes actions in order to correct it. It is also worth noting, that imperfections within measurements itself are not covered by the code. The measurement errors are accounted by performing multiple rounds of identical measurements. Finally, if the qubit and measurement error rates are below certain thresholds \cite{topological_quantum_memory, aharonov_threshold, fowler_threshold, campbell}, successful error correction is achieved.

\begin{figure}
    \centering
    \includegraphics[scale=0.55]{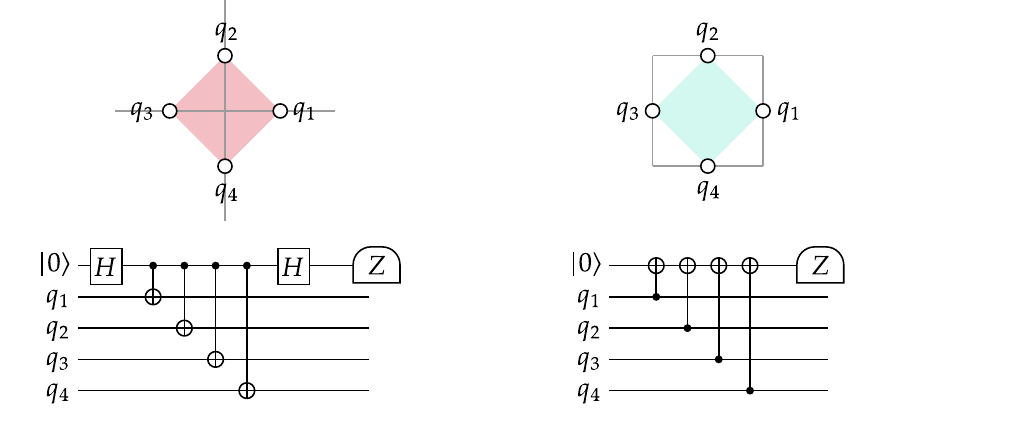}
    \caption{Quantum circuits for measurements of stabilizers of the surface code. A circuit for an X-check measurement is shown in the left, and a circuit for Z-check measurement is shown in the right.}
    \label{fig:stabilizer_checks_circuits}
\end{figure}

The stabilizer measurement can be easily transpiled into a fusion network using the gate teleportation framework. The plaquette and star stabilizer operators are equivalent to the circuits illustrated in Fig~\ref{fig:stabilizer_checks_circuits}. Four qubits $q_{i}$ are entangled using CNOT gates and a common ancillary $\ket{0}$ qubit. The ancillary qubit is measured in $Z$ basis after the CNOT gates are applied.

\begin{figure}[h]
    \centering
    \includegraphics[scale=0.5]{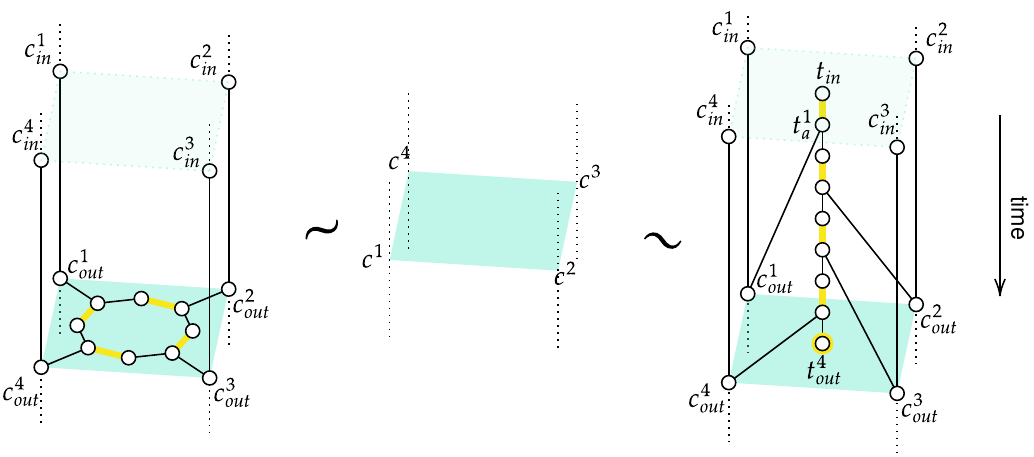}
    \caption{Fusion networks for $Z$-check measurement. In the right: implementation via the sequence of CNOT gates. In the left: cyclization scheme.}
    \label{fig:cyclization}
\end{figure}

The fusion-based equivalent of each CNOT gate requires a four-qubit linear cluster as a resource. The fusion network implementing $Z$-check measurement is presented in the right side of Fig.~\ref{fig:cyclization}. Let us denote the qubits of $j$-th resource cluster as $c_a^j$, $t_a^j$, $c_{out}^j$ and $t_{out}^j$ ($j\in\{1,\ 2,\ 3,\ 4\})$. The qubits $c_{out}^j$ carry the output state after the stabilizer construction procedure and correspond to the corner $j$ of the stabilizer square (see Fig.~\ref{fig:stabilizer_checks_circuits}). The qubits $c_a^j$ are fused with the initial qubits $q_{j}$. For $j>1$ $t_a^j$ are fused with $t_{out}^{j-1}$. Finally qubit $t_a^1$ is fused with the ancillary qubit $t_a^0$, which is initialized in state $\ket{0}$, and $t_{out}^4$ is measured in the $Z$-basis. One can notice, that the check measurement can be implemented not only for the chosen circular enumeration of the stabilizer square corners as in Fig.~\ref{fig:stabilizer_checks_circuits}, but also for arbitrary permutation of corner indices. The same procedure can be derived for $X$-check operators.

\begin{figure}[h]
    \centering
    \includegraphics[scale=0.5]{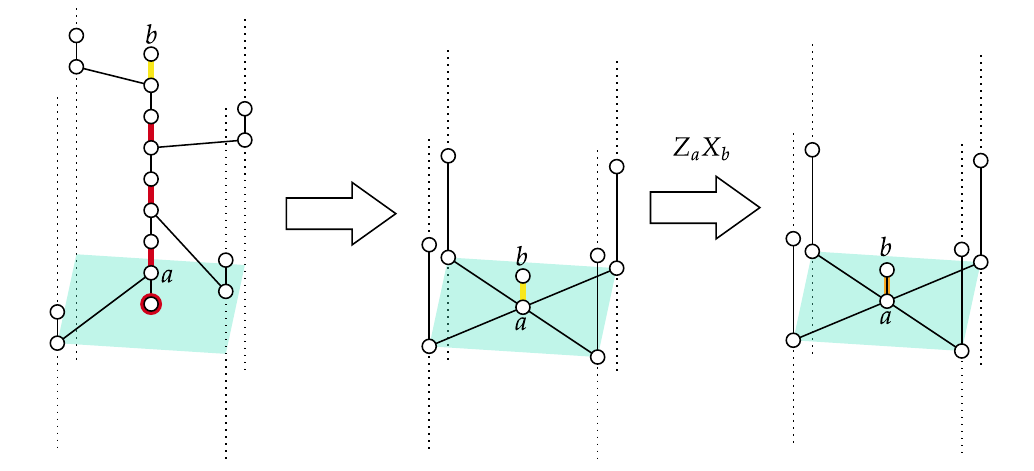}
    \caption{Three stages of fusions of qubits, that correspond to ancilla qubit of $Z$-check measurement. In the left part the first stage fusions are colored as red. In the middle part there is the result of "red" fusions. The configuration of measurements and state after measurement of $Z_aX_B$ is presented in the right part. Orange line connecting qubits $a$ and $b$ corresponds to $X_aZ_b$-measurement.}
    \label{fig:noncyclization_stages}
\end{figure}

\begin{figure}[h]
    \centering
    \includegraphics[scale=0.5]{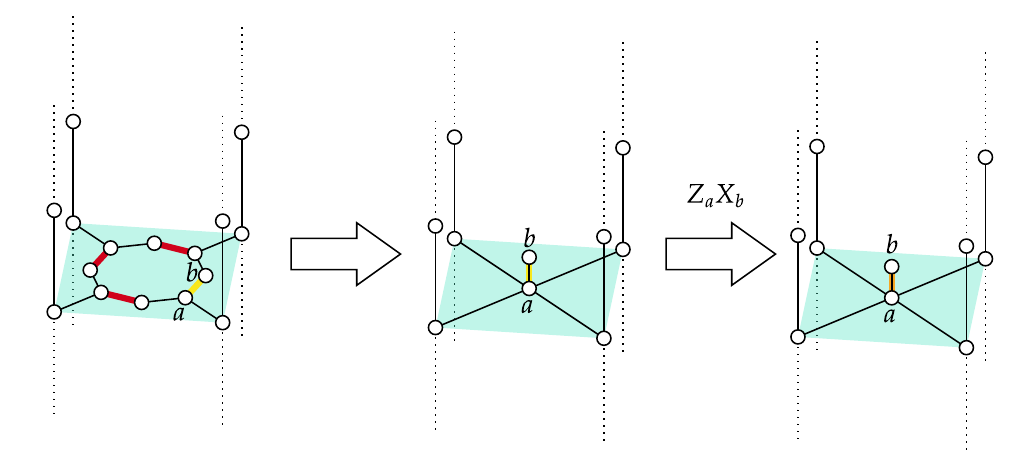}
    \caption{Three stages of fusion of the central qubits of the cyclization scheme. The left part describes the fusions colored in red which are performed at the first stage. The middle part illustrates the result of these fusions. The configuration of measurements and state after the $Z_aX_B$ measurement is presented in the right part. Orange line connecting qubits $a$ and $b$ corresponds to $X_aZ_b$-measurement.}
    \label{fig:cyclization_stages}
\end{figure}

Thus, the transpilation of the circuits shown in Fig.~\ref{fig:stabilizer_checks_circuits} results in the fusion network of 4 four-qubit linear clusters and a single qubit. The network involves 4 two-qubit measurements and one single-qubit measurement. We transform this fusion network to an equivalent one which includes only four-qubit cluster states and fusion measurements. We apply a strategy that we call \textit{cyclization}. Instead of the $Z$-measurement of the qubit $t_{out}^4$ in the $Z$-check operator network we remove ancillary qubit state $\ket{0}$ and fuse the $t_{out}^4$ and $t_{a}^1$ qubits which completes a cycle of fusion measurements inside the network. The sequence is illustrated in the left part of Fig.~\ref{fig:cyclization}. The cyclization procedure provides the identical stabiliser measurement. The $X$-check operator adopts similar cyclization. It is easier to see the equivalence of cyclization and non-cylization schemes, if we fuse the central qubits of fusion network in three stages. For the scheme without cyclization, in the first stage we fuse qubits $t_{out}^j$ for $j>1$. The procedure is illustrated in Fig.~\ref{fig:noncyclization_stages}. In the left part of the picture, the performed fusions are colored as red, while in the middle, we have the network that consists of two states. The first one is the cluster of nine qubits and the second is a single qubit $b$, which is in the state $\ket{0}$ and is to be fused with qubit $a$. Note, $Z_aX_b$ is one of the observables of the remaining fusion of qubits $a$ and $b$. Measuring this operator results in a $10$ qubit cluster. This cluster is identical to the one produced from the cyclization scheme, when three out of four central fusions are performed, see the left and the middle parts of Fig.~\ref{fig:cyclization_stages}. The operator $Z_aX_b$ is a stabilizer of this cluster. Therefore, its measurement does not alter the state. Thus, as shown in the right parts of Fig.~\ref{fig:noncyclization_stages} and Fig.~\ref{fig:cyclization_stages}, both cyclization and non-cylization schemes lead to the same configuration of states and measurements. The final step involves the measurement of $X_aZ_b$ in the 10-qubit cluster.

Remarkably, the cyclization procedure not only simplifies the fusion network of stabilizer measurement, but also gives rise to additional check operator. Indeed, the product of the measured operators $X_{t_{out}^{j}}Z_{t_a^{j+1}}\ \ (j=1..3)$ and $X_{t_{out}^4}Z_{t_a^1}$ is also a stabilizer of the state of the involved 4-qubit linear clusters and qubit $t_{out}^0$. This means that if there are no measurement errors, the product of outcomes of $X_{t_{out}^{j}}Z_{t_a^{j+1}}\ \ (j=1..3)$ and $X_{t_{out}^4}Z_{t_a^1}$ must be equal to $1$. However, as we will see in the next section, the cyclization also leads to the emergence of multi-edges and that worsens the fault-tolerant properties of the resulting syndrome graph.

\begin{figure*}
    \centering
    \includegraphics[scale=1.0]{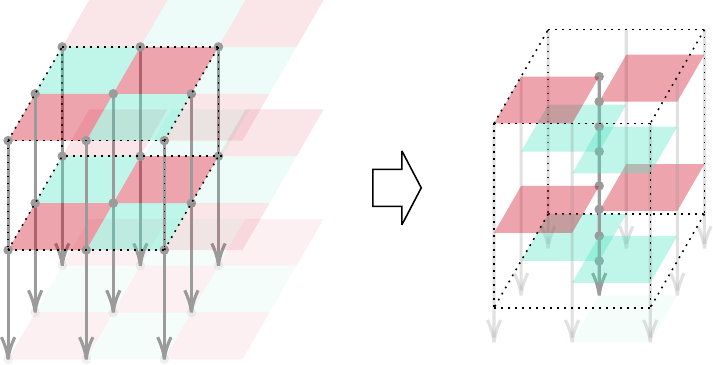}
    \caption{Time and space ordering of stabilizer measurements. In the left: initial quantum circuit, where all operations of the same round of stabilizer measurements are performed simultaneously. In the right: separation of $X$-check and $Z$-check measurements in time. Grey lines correspond to the register qubits timelines, red and blue squares are associated with $X$-check and $Z$-check stabilizer measurements. The time axis points downwards. }
    \label{fig:ordering}
\end{figure*}

We assemble the fusion network corresponding to the surface code by interconnecting the fusion-based implementations of $X$-check and $Z$-check operators. This can be performed using multiple strategies and we outline a specific one below. We build a fusion network that implements consecutive rounds of stabilizers measurement. Let us start from rearranging the qubit register on a flat 2D surface. The corresponding quantum circuit is presented in the left side of Fig.~\ref{fig:ordering}. There, every colored square denotes a stabilizer measurement operation. The grey lines show the timelines for the register qubits propagation. Thus, we have a 3D structure, where one axis corresponds to time direction. In the presented quantum circuit all stabilizers belonging to the same round are measured simultaneously. However, every qubit from the register participates in four different stabilizer measurements. Every stabilizer measurement operation consumes a register qubit as input, fuses it with an ancillary qubit and teleports its state to the output qubit, that can be used in the next measurement. At the same time, the register qubit can not survive the stabilizer check measurement. Therefore, one need to separate the performed measurement operations in space and time and fix the corresponding ordering.

Since the graph $G$ that defines the surface code is bipartite, we can divide its set of vertices $V(G)$ into two groups $V_1(G)$ and $V_2(G)$. These groups are defined such that there are no pairs of vertices connected by an edge within each group. A similar partition can be introduced for graph $G^*$. Every register qubit corresponds to an edge in $G$, that connects vertices $v_1$ and $v_2$ from subsets $V_1(G)$ and $V_2(G)$ respectively. The dual counterpart of this edge also connects $u_1$ and $u_2$ from subsets $V_1(G^*)$ and $V_2(G^*)$. Thus, the considered qubit participates in measurement of two $X$-check stabilizers corresponding to vertices $v_1$ and $v_2$ of $G$, and in measurement of two $Z$-check stabilizers corresponding to plaquettes $u_1$ and $u_2$ of $G$. Separating this four measurements we obtain the final time ordering, that is depicted in the right part of Fig.~\ref{fig:ordering}. For every stabilizers measurement round, we start from the measurement of $X$-checks corresponding to vertices from $V_1(G)$. Then, we perform measurements of $X$-checks of $V_2(G)$ vertices. After that, the measurement of $Z$-checks is made, which is also divided into two stages: at first $V_2(G^*)$ stabilizers are measured, followed by $V_1(G^*)$ measurements.

\begin{figure}
    \centering
    \includegraphics[scale=0.5]{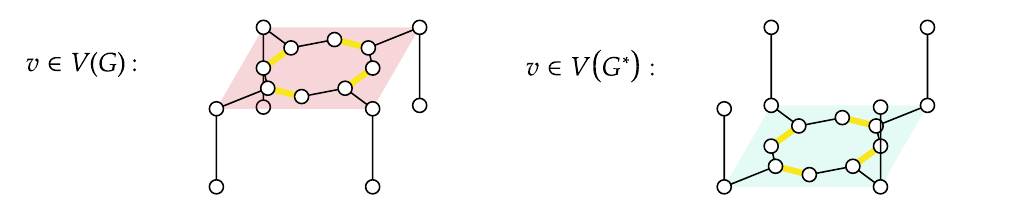}
    \caption{Fusion networks for the measurements of surface code stabilizers. Red and blue squares indicates the fusion configuration for $X$-check and $Z$-check operators relatively.}
    \label{fig:stabilizer_fn}
\end{figure}

Finally, we present the stabilizer check measurements as fusion networks from Fig.~\ref{fig:stabilizer_fn} and obtain the fusion network for the foliated surface code based on the $4$-qubit linear cluster states. Let us refer to this construction as the four-qubit FTFN.

\begin{figure}
    \centering
    \includegraphics[width=0.85\linewidth]{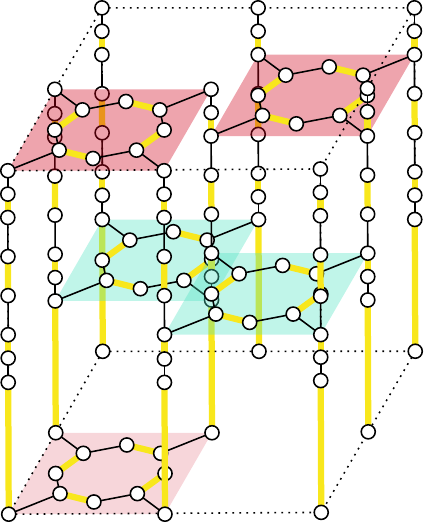}
    \caption{The unit cell of the four-qubit FTFN. Yellow edges correspond to fusions.}
    \label{fig:4qubit_model_cell}
\end{figure}


\begin{figure*}
    \centering
    \includegraphics[width=0.85\linewidth]{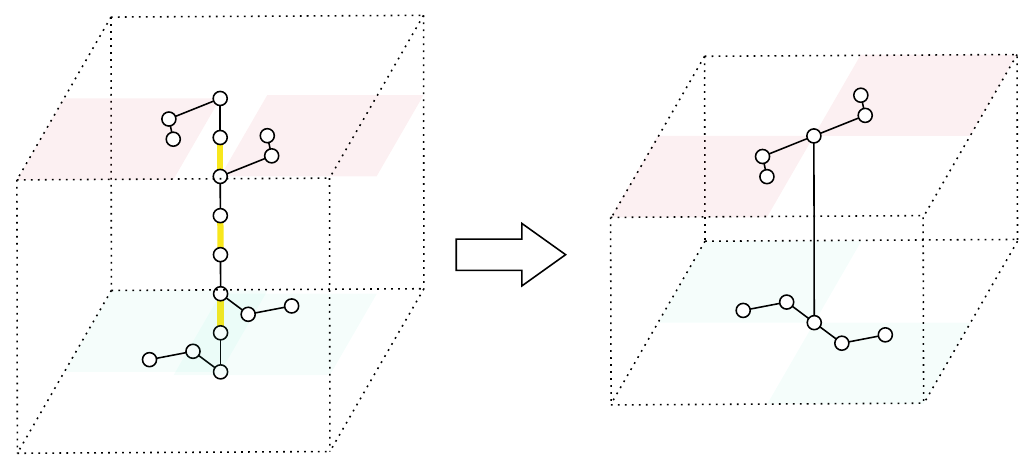}
    \caption{Construction the 10-qubit resource state from the 4-qubit linear clusters.}
    \label{fig:4q_to_10q_res}
\end{figure*}

\begin{figure}
    \includegraphics[width=0.8\linewidth]{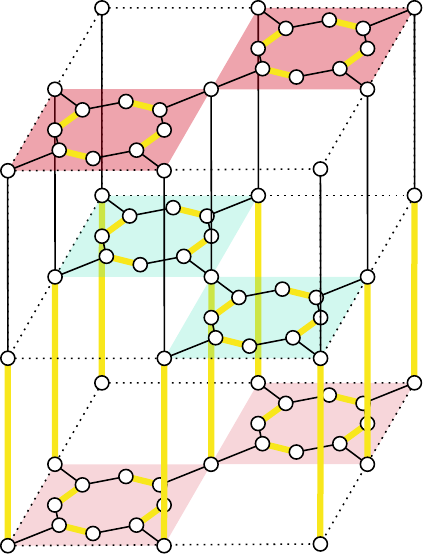}
    \caption{The unit cell of the ten-qubit model FTFN. Yellow edges correspond to fusions.}
    \label{fig:10qubit_model_cell}
\end{figure}

The four-qubit model naturally produces other FTFNs if some of the resource states are fused together. For instance, we can fuse all four-qubit clusters along the line corresponding to the teleportation of a physical qubit within a round of a check measurement, see Fig.~\ref{fig:4q_to_10q_res}. In other words, we perform all fusions that are not involved in the cyclization and do not outspread between different check measurement rounds. Then, we get the ten-qubit FTFN with ten-qubit resource states, which we refer to as the ten-qubit model. The ten-qubit FTFN structure is outlined in Fig.~\ref{fig:10qubit_model_cell}.

\begin{figure*}
    \centering
    \includegraphics[scale=0.8]{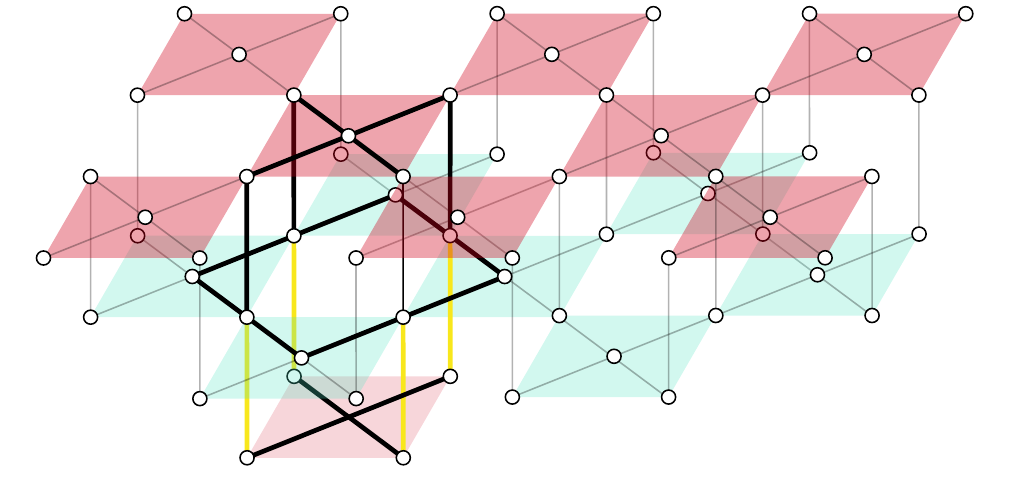}
    \caption{The Raussendorf equivalent FTFN. Yellow edges correspond to fusions. The unit cell is highlighted.}
    \label{fig:raussendorf_equivalent}
\end{figure*}

We get one more FTFN if we modify the four-qubit FTFN without employing the cyclization. For every square from Fig.~\ref{fig:10qubit_model_cell} we replace the set of qubit participating in cyclization with a single qubit. This qubit is connected by edges with all qubits located on the vertices of the square. We perform all fusions that connect clusters within a single round of check measurements and finish with a layered structure presented in Fig.~\ref{fig:raussendorf_equivalent}.

If we unite all the graphs of the resource states in the layered structure by connecting all pairs of fusing qubits by $CZ$ edges, we obtain the Raussendorf cluster state \cite{raussendorf_fault_tolerant_MBQC}.The Raussendorf lattice is a bipartite graph where the vertices of the first part, called \textit{primal}, can be associated with the faces of a cubic lattice $Q$. The vertices of the second part, called \textit{dual}, are located on the edges of the same cubic lattice $Q$. For every face, the edges of the Raussendorf lattice connect the primal vertex with dual vertices located on the border of that face. It is worth noting that primal vertices transform into dual ones under translation by half period of $Q$ in all three directions. This translation maps the Raussendorf lattice onto itself and the graph $Q$ to its dual $Q^*$, which is obviously cubic. The presented cluster state is used in the Raussendorf MBQC model, which describes an architecture for fault-tolerant quantum computation.

The Raussendorf model admits an FBQC description, if all fusions are single-qubit measurements in $X$-basis, i.e. the fusion group is generated by operators $X_q$ for all measured qubits. The group $\mathcal{R}$ consists of generators of the Raussendorf cluster state, i.e.

\begin{equation}
\begin{array}{c}
    g^p_v = X_v\prod_{u\in B_Q(v)}Z_u,\quad v\in faces(Q),\\
    g^d_v = X_v\prod_{u\in B_{Q^*}(v)}Z_u,\quad v\in faces(Q^*),
\end{array}
\end{equation}
where $faces(Q)$ is a set of all faces of the lattice $Q$ and $B_Q(v)$ is a set of edges borders of the face $v$, the $Q^*$ is a graph dual to $Q$. One may see that if $M\subset faces(Q)$ the expression $\prod_{v\in M}g^p_v = \prod_{v\in M}X_v\prod_{u\in B_Q(M)}Z_u$ is valid. Here the $B_Q(M)$ is a border of a disjoint union of faces from the set $M$. The same property holds for operators $g^d_v$. For the MBQC model based on the Raussendorf lattice, one implies that every element of the check operator group $C=R\cap F$ is a product of single-qubit $X$ operators. Therefore, the elements from $C$ correspond to the surfaces without a boundary in the lattices $Q$ and $Q^*$. An arbitrary volume in the lattices $Q$ and $Q^*$ is a union of corresponding elementary cells. The set of all faces enclosing the volume does not have a border. Therefore, the elements from $R$ that correspond to elementary cells of the lattices $Q$ and $Q^*$ are the generators of the group $C$.

\begin{equation}
\begin{array}{cc}
         c_{cell} = \prod_{f\in cell}X_f,\\
         \text{for}\ cell\in cells(Q)\ \text{or}\ cell\in cells(Q^*).
\end{array}
\end{equation}

\begin{figure*}
\centering
    \includegraphics[width=0.8\linewidth]{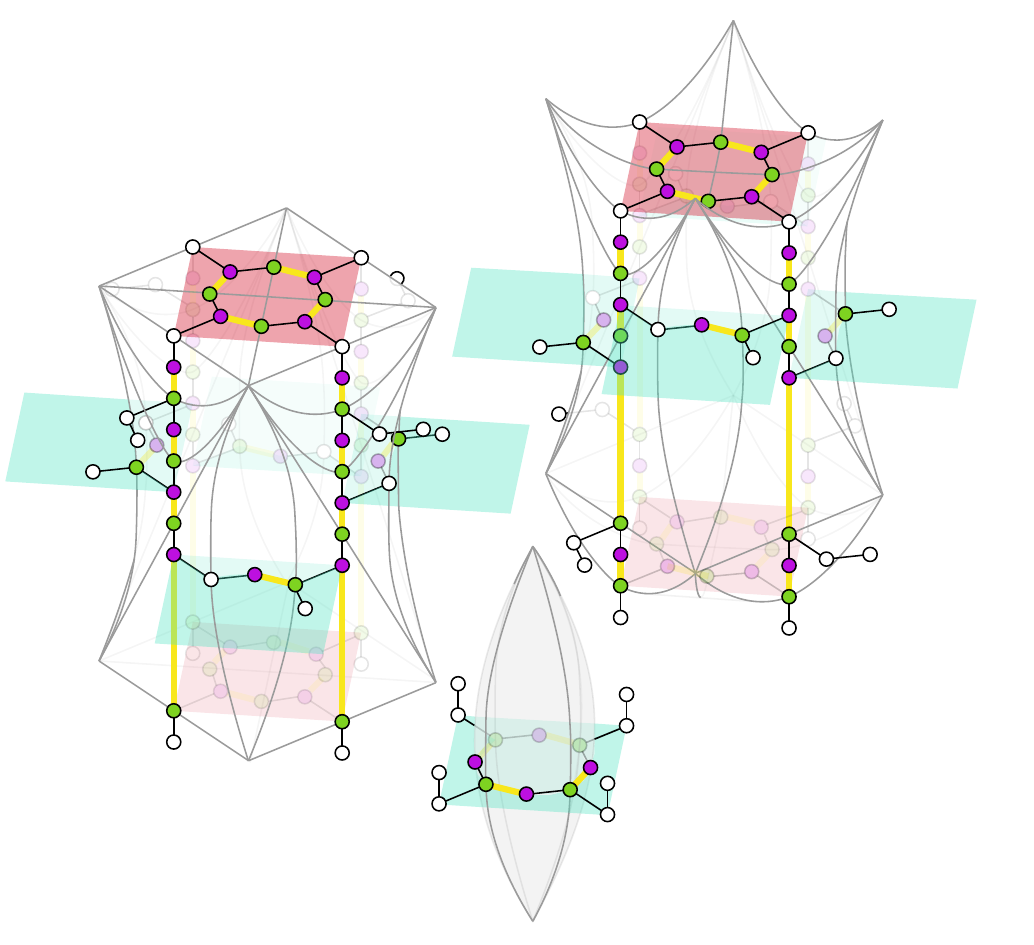}
    \caption{Construction of the dual syndrome graph $L$ for the $4$-qubit model. Its edges are depicted as grey lines. Violet circles corresponds to primal qubits, white and green ones denote dual qubits. The colors shows the Pauli operators whose products are elements of $C=R\cap F$: violet for $X$, green for $Z$ and white for identity. Yellow edges connect the fusing qubits. Grey lines are edges of the lattice $L$. There are two types of the $28$-faced cells in the left and in the right of the picture and a $4$-faced cell in the center.}
    \label{fig:4q_syndrome_cell}
\end{figure*}

It is convenient to study the fault-tolerant properties of the FBQC models introduced in the present paper starting from the Raussedorf model. First, let us note that the $\langle X_{q_1}Z_{q_2},\ Z_{q_1}X_{q_2}\rangle$ fusion is equivalent to performing a $CZ(q_1,\ q_2)$ operation followed by destructive measurements $X_{q_1}$ and $X_{q_2}$. The constructed layered FTFN (see Fig.~\ref{fig:raussendorf_equivalent}) is equivalent to the Raussendorf model if we apply $X$ measurement to every qubit of the Raussendorf state. This equivalence also means that the syndrome graphs of the models are identical up to the operator substitutions $X_f\to X_fZ_e$ and $X_e\to Z_fX_e$ (for appropriate $f$ and $e$). On the left side of Fig.~\ref{fig:4q_syndrome_cell}, we depict a part of the layered FTFN that corresponds to an elementary cell of the cubic lattice $Q$. Here, we associate the horizontal faces of the constructed graph $Q$ with $X$-check stabilizers, and the horizontal faces of $Q^*$ with $Z$-check measurements. The produced syndrome graph has two components and the depicted cell corresponds to a vertex in one of them. Thus, graphs $L$ and $L^*$ are actually constitute the syndrome graph of the layered FTFN. For each component the horizontal edges correspond to the fusion gates and vertical ones correspond to $X$ measurements of ancillary qubits.

\begin{figure*}
\centering
    \includegraphics[width=0.8\linewidth]{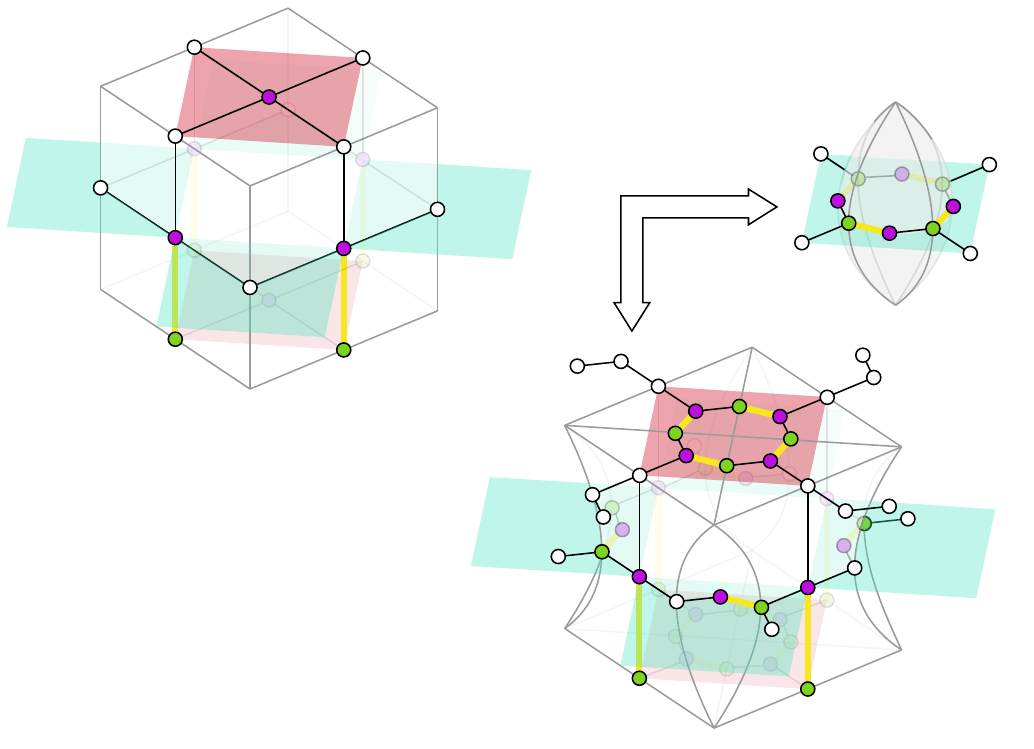}
    \caption{Transformation from the Raussendorf qubic cell $Q$ (see the left part of the picture) to the lattice $L$ (in the right part), which is dual to the syndrome graph of the $10$-qubit model. Every primal qubit (violet) corresponds to a face of $L$, dual qubit (white and green) correspond to edges of $L$ (grey lines). The colors show the Pauli operators the products of which are elements of $C=R\cap F$: violet for $X$, green for $Z$ and white for identity. Yellow edges connect the fusing qubits. Grey lines are edges of the lattice $L$. There are two types of cell in $L$. The first one is a $16$-faced cell in the bottom right part of the picture and the second one is a $4$-faced cell in the top right.}
    \label{fig:10q_syndrome_cell}
\end{figure*}

\begin{figure}
\centering
    \includegraphics[width=0.8\linewidth]{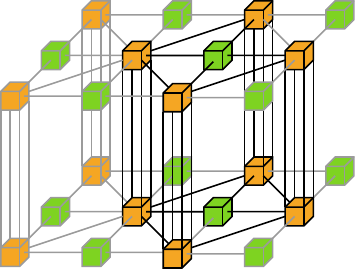}
    \caption{Two layers of the syndrome graph of the $10$-qubit model with a unit cell highlighted. This graph is dual to the lattice $L$ and has the same structure. The orange vertices correspond to $16$-faced cells and each one of them has $8$ incident edges and $2$ incident edges of multiplicity $4$. Each green vertex has $4$ incident edges and correspond to $4$-faced cells of $L$.}
    \label{fig:10q_syndrome_graph}
\end{figure}

Now we move to description of the syndrome graph for the ten-qubit FTFN. Here, we also use equivalence between the initial FBQC FTFN and MBQC model that is based on cluster state produced from the initial FTFN by connecting fusing pairs by edges. The qubits are divided into primal and dual sets in an alternating order, i.e. no primal qubit has a primal neighbour. This cluster state can be obtained from the Raussendorf lattice by substituting all the qubits corresponding to horizontal faces of cubic lattices $Q$ and $Q^*$ by cycles of length $8$, see Fig.~\ref{fig:10q_syndrome_cell}. We construct a lattice $L$ by associating each primal qubit with a face and each dual qubit with an edge of $L$. Unlike the graph $Q$ here we have two kinds of cells. The first one is produced by ``slicing`` corners of the initial cubic cell and dividing its horizontal faces into four parts. The second type of cells corresponds to the inserted $8$-cycles, see Fig.~\ref{fig:10q_syndrome_cell}. The cluster state corresponding to the ten-qubit FTFN is self dual, therefore graph $L^*$ has the same structure as $L$. Also, each cell of these lattices corresponds to a vertex of the syndrome graph of the ten-qubit FTFN. Thus, $L$ and $L^*$ are two components of the syndrome graph. In Fig.~\ref{fig:10q_syndrome_graph} we can see the part of its structure with a unit cell highlighted.

\begin{figure}
\centering
    \includegraphics[width=0.8\linewidth]{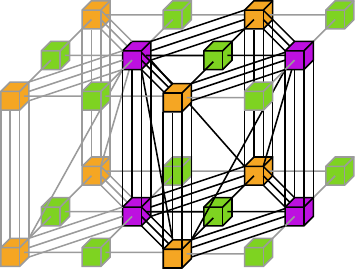}
    \caption{Two layers of the syndrome graph of the $4$-qubit model with highlighted unit cell. The orange and violet vertices have $8$ incident edges, $4$ incident edges of multiplicity $3$, and $2$ incident edges of multiplicity $4$. The green vertices have $4$ incident edges.}
    \label{fig:4q_syndrome_graph}
\end{figure}

The construction of the syndrome graph of the four-qubit FTFN can be accomplished by modification of the cells of graphs $L$ and $L^*$ that is defined above for the ten-qubit FTFN. Let us concentrate on $L$. We obtain different results after modification of cells corresponding to $v_1\in V_1(G)$ and $v_2\in V_2(G)$, because $v_1$ connects with the fusion structures implementing the previous round of the stabilizer measurements and $v_2$ does not. Finally, we have three types of cells. The first one with four faces as a border corresponds to the cyclization scheme and is inherited from graph $L$ without modifications. Each one from the other two types is produced from the cell with sliced corners of $L$ by dividing the side faces into four parts. Therefore, these cells have $28$ faces and correspond to the vertices from either $V_1(G)$ or $V_2(G)$, see Fig.~\ref{fig:4q_syndrome_cell}. Let us call them $V_1$-type and $V_2$-type cells. The three faces of the divided side face connect one $V_1$-type cell and one $V_2$-type cell located in the same layer. The fourth face of the side face's division connects a $V_1$-type cell with a $V_2$-type cell from the previous round of stabilizer measurements. The structure of the syndrome graph with highlighted unit cell is shown in Fig.~\ref{fig:4q_syndrome_graph}. The rectangular blocks depict vertices of the syndrome graph. The green blocks correspond to the inherited $4$-faces cells. The orange and purple blocks correspond to $V_1$-type and $V_2$-type cells, respectively. One can see the triple edges connecting the orange and purple blocks within one layer and singular edges connecting the orange and purple blocks from neighboring layers. These edges correspond to the division of the side face of the sliced cell.

\section{Noise tolerance}
\label{sec: noise tolerance}

The proposed four-qubit and ten-qubit FTFNs can serve as bases for the fault-tolerant quantum computing models. The check operator group in both models is not empty and generates a syndrome graph. In this section, we study the performance of the models by simulating the noise process and estimate the thresholds for error and erasure rates.

We follow the original FBQC guideline for the fault tolerance analysis \cite{FBQC_1} and consider two types of noise effects: an error (or fusion outcome flip) and an erasure of a fusion outcome. The first one originates from the imperfections in the resource state generation and measurements in fusion gates. These processes lead to parasitic flips in the fusion measurement outcomes. We describe this type of noise effect as the Pauli errors occurring on qubits according to independent identical distributions. Additionally, assuming that the probability of a bit flip is equal to the probability of a phase flip, one can see that the probabilities of measurement outcome flips in fusion gates are the same. We denote this parameter as $p_{error}$.

An ideal fusion gate acting on qubits $a$ and $b$ outputs two outcomes for the $X_aZ_b$ and $Z_aX_b$ operators. However, it may occur that one or both outcomes are lost. In the context of a linear optics setup, this can be caused by photon loss and by the intrinsic non-determinism of the fusion gates \cite{LOCC,fusion_types}. Photon loss occurs, when we do not observe two detected photons. In the case, the outcomes of both observables are lost. At the same time, we may lose outcomes, even if we detected all measured photons. In linear optical implementation of fusion of dual-rail qubits, not all detection patterns correspond to genuine projection onto the mutual eigenstate of the operators $X_aZ_b$ and $Z_aX_b$. For the simplest linear optical circuits, in half of the cases associated with the failure of the fusion, there will be projection onto eigenstate of operators $X_a$ and $Z_b$. However, one can use more complicated linear optical schemes and additional resources in order to decrease the probability of obtaining the failure patterns \cite{boosted_fusion}. The measurement of $X_a$ and $Z_b$ is equivalent to the measurement of $X_aZ_b$ and $Z_b$. Thus, in the case of failure, we still obtain outcomes of one of the measured observables. One can also choose the linear optical fusion scheme, where in the case of failure, the observable $X_aZ_b$ is erased and outcome of $Z_aX_b$ is obtained. The problem of setting the fusions for fusion networks is known as {\it bias arrangement}, and we refer to \cite{FBQC_increasing_error_tolerance} for more details. In the present paper, we suggest that for each fusion the particular configuration is set randomly.  In the case of failure, we lose the outcome of either $X_aZ_b$ or $Z_aX_b$ with equal probability. Taking into account both causes of erasure, we suppose that each outcome can be lost with a probability of $p_{erasure}$. We refer to it as the erasure probability.

We perform Monte-Carlo simulations to to estimate the fault tolerant region in the space of parameters $p_{error}$ and $p_{erasure}$, i.e. all values of  $p_{error}$ and $p_{erasure}$ for which one can successfully run error correction procedures. The details about using error model and simulation are presented in App.~\ref{sec: simulation details}. The results for both the four-qubit and the ten-qubit models are plotted in Fig.~\ref{fig:thresholds}.

In the absence of the error noise the erasure threshold for the ten-qubit model reaches $12.8\%$, which is higher than the hexagon model erasure threshold reported in the original FBQC paper \cite{FBQC_1} ($11.98\%$). This model also shows the better result for error threshold: $1.33\%$ vs $1.08\%$. We also compare our results with performance of models based on the foliated Floquet color code \cite{floquet_QC}. The ten-qubit model does not achieve threshold values of $\infty$-chain model ($13.2\%$ for the erasure rate and $1.5\%$ for the error rate), but it outperforms the $14$-chain model.

The $4$-qubit model does not match the performance of the ten-qubit model. In this case, the erasure threshold is equal to $6.4\%$, and the error threshold is equal to $0.58\%$. As one can see, these values are lower than those reported for the star resource state model \cite{FBQC_1} ($6.9\%$ and $0.75\%$ correspondingly).

\section{Conclusion}
\label{sec: conclusion}
We presented the constructive algorithm to build fault-tolerant fusion networks using the fusion-based equivalent of the gate teleportation circuits. We test the proposed principle on the foliated surface code model and output the four-qubit and ten-qubit fault-tolerant fusion networks. Lastly we conduct the numerical analysis and estimate the $p_{error}$ and $p_{erasure}$ thresholds. Though, we were mainly focused on demonstration of the presented approach, the constructed FTFNs show performance comparable with the previously reported fusion networks. Moreover, the ten-qubit FTFN had the one of the highest error and erasure thresholds among models investigated in \cite{FBQC_1,floquet_QC} (it loses only to encoded version of the hexagon model from \cite{FBQC_1} and to the $\infty$-chains model from \cite{floquet_QC}). However, the ten-qubit FTFN has more complicated structure
of fusion graph and resource states than the hexagon model from \cite{FBQC_1}. 

The fusion-based quantum computation model can be applied using any physical platform for quantum computing. However, it was introduced mainly due to the necessity to tackle the intrinsic issues of the linear optical quantum computer architecture such as high qubit loss rate and non-deterministic entangling operations. In linear optics, the problem of generation of entangled photonic states is vital. One can see that the fusion networks employing complex cluster states are more robust to noise. However, it is not clear how to produce these resources effectively. One can use cumbersome multiplexing schemes for their generation, which rely on more complex and deep linear optical networks, but then the higher level of photon loss is introduced.
The hardware resources required for four-qubit and ten-qubit state generation still have to be accurately estimated, but it is clear that the ten-qubit state is substantially harder to generate than the four-qubit state, especially in the case of linear optics. Linear optical circuits for standard entangled state generation (i.e., Bell state, GHZ state, and some other types) have been well studied \cite{KLM, knill, pan-rudolph, fldzhyan-bell, uskov2015, wehner2019, varnava2008}, but the generic case of finding an optimal circuit for entangled state generation has no established solution. Several optimization-based algorithms have been reported in \cite{uskov2009, eisert2004,  gubarev} and yet none of them can guarantee the optimal design of the circuit in terms of the success probability of state generation. In any case, the larger number of photons in the entangled state leads to an exponentially smaller success probability.

We believe that our results will motivate further research of fault-tolerant fusion networks since we provide the guidelines on how to build such networks using the existing error-correcting codes. Our construction may also be used for direct transpilation of quantum circuits into fusion networks facilitating the wider adoption of these new computational model.

\section{Acknowledgements}
The authors acknowledge support by Rosatom in the framework of the Roadmap for Quantum computing (Contract No. 868-1.3-15/15-2021 dated October 5, 2021 and Contract No.P2154 dated November 24, 2021). I. D. acknowledges support from Russian Science Foundation, grant No. grant 22-12-00353-P.

\bibliographystyle{quantum}

\begin{thebibliography}{9}

\bibitem{elementary_gates}
A. Barenco, C. H. Bennett, R. Cleve, D. P. DiVincenzo, N. Margolus, P. Shor, T. Sleator, J. A. Smolin, H. Weinfurter. Elementary gates for quantum computation. \href{https://doi.org/10.1103/PhysRevA.52.3457}{Phys. Rev. A 52, 3457--3467} (1995).

\bibitem{kitaev_solovey}
A. Y. Kitaev. Quantum computations: algorithms and error correction. \href{https://doi.org/10.1070/RM1997v052n06ABEH002155}{Russ. Math. Surv. 52, 1191–1249} (1997).

\bibitem{nielsen_chang}
M. A. Nielsen, I. Chuang. Quantum Computation and Quantum Information. Cambridge (2010). Cambridge University Press.

\bibitem{annealing}
A. B. Finnila, M. A. Gomez, C. Sebenik, C. Stenson, J. D. Doll. Quantum annealing: A new method for minimizing multidimensional functions. \href{https://doi.org/10.1016/0009-2614(94)00117-0}{Chemical Physics Letters 219, 343–348} (1994).

\bibitem{adiabatic_computation}
E. Farhi, J. Goldstone, S. Gutmann, M. Sipser. Quantum Computation by Adiabatic Evolution. \href{https://doi.org/10.48550/arXiv.quant-ph/0001106}{arXiv:quant-ph/0001106} (2000).

\bibitem{anyons}
A. Y. Kitaev. Fault-tolerant quantum computation by anyons. \href{https://doi.org/10.1016/S0003-4916(02)00018-0}{Annals of Physics 303, 2–30} (2003), \href{https://doi.org/10.48550/arXiv.quant-ph/9707021}{arXiv:quant-ph/9707021} (1997).

\bibitem{topological_computation}
C. Nayak, S. H. Simon, A. Stern, M. Freedman, S. Das Sarma. Non-Abelian anyons and topological quantum computation. \href{https://doi.org/10.1103/RevModPhys.80.1083}{Rev. Mod. Phys. 80, 1083–1159} (2008).

\bibitem{raussendorf_one-way}
R. Raussendorf, H. J. Briegel. A One-Way Quantum Computer. \href{https://doi.org/10.1103/PhysRevLett.86.5188}{Phys. Rev. Lett. 86, 5188–5191} (2001).

\bibitem{raussendorf_MBQC}
R. Raussendorf, D. E. Browne, H. J. Briegel. Measurement-based quantum computation on cluster states. \href{https://doi.org/10.1103/PhysRevA.68.022312}{Phys. Rev. A 68, 022312} (2003).

\bibitem{FBQC_1}
S. Bartolucci, P. Birchall, H. Bombín, H. Cable, C. Dawson, M. Gimeno-Segovia, E. Johnston, K. Kieling, N. Nickerson, M. Pant, F. Pastawski, T. Rudolph, C. Sparrow. Fusion-based quantum computation. \href{https://doi.org/10.1038/s41467-023-36493-1}{Nat Commun 14, 912} (2023).

\bibitem{FBQC_interleaving}
H. Bombin, I. H. Kim, D. Litinski, N. Nickerson, M. Pant, F. Pastawski, S. Roberts, T. Rudolph. Interleaving: Modular architectures for fault-tolerant photonic quantum computing. \href{https://doi.org/10.48550/arXiv.2103.08612}{arXiv:quant-ph/2103.08612} (2021).

\bibitem{FBQC_logical_blocks}
H. Bombin, C. Dawson, R. V. Mishmash, N. Nickerson, F. Pastawski, S. Roberts. Logical blocks for fault-tolerant topological quantum computation. \href{https://doi.org/10.48550/arXiv.2112.12160}{arXiv:quant-ph/2112.12160} (2021).

\bibitem{FBQC_unifying_with_ZX}
H. Bombin, D. Litinski, N. Nickerson, F. Pastawski, S. Roberts. Unifying flavors of fault tolerance with the ZX calculus. \href{https://doi.org/10.48550/arXiv.2303.08829}{arXiv:quant-ph/2303.08829} (2023)

\bibitem{FBQC_increasing_error_tolerance}
H. Bombín, C. Dawson, N. Nickerson, M. Pant, J. Sullivan. Increasing error tolerance in quantum computers with dynamic bias arrangement. \href{https://doi.org/10.48550/arXiv.2303.16122}{arXiv:quant-ph/2303.16122} (2023).

\bibitem{fusion_types}
D. E. Browne, T. Rudolph. Resource-efficient linear optical quantum computation. \href{https://doi.org/10.1103/PhysRevLett.95.010501}{Phys. Rev. Lett. 95, 010501} (2005).

\bibitem{qudits_computation}
J. Joo, P. L. Knight, J. L. O’Brien, T. Rudolph. One-way quantum computation with four-dimensional photonic qudits. \href{https://doi.org/10.1103/PhysRevA.76.052326}{Phys. Rev. A 76, 052326} (2007).

\bibitem{percolation_mbqc}
K. Kieling, T. Rudolph, J. Eisert. Percolation, renormalization, and quantum computing with non-deterministic gates. \href{https://doi.org/10.1103/PhysRevLett.99.130501}{Phys. Rev. Lett. 99, 130501} (2007).

\bibitem{gimeno-segovia_3ghz}
M. Gimeno-Segovia, P. Shadbolt, D. E. Browne, T. Rudolph. From three-photon GHZ states to ballistic universal quantum computation. \href{https://doi.org/10.1103/PhysRevLett.115.020502}{Phys. Rev. Lett. 115, 020502} (2015). 

\bibitem{omkar1}
S. Omkar, S.-H. Lee, Y. S. Teo, S.-W.Lee, S.-W., H. Jeong. All-photonic architectural roadmap for scalable quantum computing using Greenberger-Horne-Zeilinger states. \href{https://doi.org/10.1103/PRXQuantum.3.030309}{PRX Quantum 3, 030309} (2022).

\bibitem{omkar2}
S.-H. Lee, S. Omkar, Y. S. Teo, H. Jeong. Parity-encoding-based quantum computing with Bayesian error tracking. \href{https://doi.org/10.1038/s41534-023-00705-9}{npj Quantum Inf 9, 39} (2023).

\bibitem{zhang_compilation}
H. Zhang, A. Wu, Y. Wang, G. Li, H. Shapourian, A. Shabani, Y. Ding. A Compilation Framework for Photonic One-Way Quantum Computation. \href{https://doi.org/10.1145/3579371.3589047}{Proceedings of the 50th Annual International Symposium on Computer Architecture} (2023).

\bibitem{cluster_states}
H. J. Briegel, R. Raussendorf. Persistent Entanglement in Arrays of Interacting Particles. \href{https://doi.org/10.1103/PhysRevLett.86.910}{Phys. Rev. Lett. 86, 910–913} (2001).

\bibitem{raussendorf_fault_tolerant_MBQC}
R. Raussendorf, J. Harrington, K. Goyal. A fault-tolerant one-way quantum computer. \href{https://doi.org/10.1016/j.aop.2006.01.012}{Annals of Physics 321, 2242–2270} (2006).

\bibitem{brown_foliation}
B. J. Brown, S. Roberts. Universal fault-tolerant measurement-based quantum computation. \href{https://doi.org/10.1103/PhysRevResearch.2.033305}{Phys. Rev. Res. 2, 033305} (2020).

\bibitem{stace_foliation}
A. Bolt, G. Duclos-Cianci, D. Poulin, T. M. Stace. Foliated Quantum Error-Correcting Codes. \href{https://doi.org/10.1103/PhysRevLett.117.070501}{Phys. Rev. Lett. 117, 070501} (2016).

\bibitem{beyond_foliation}
N. Nickerson, H. Bombín. Measurement based fault tolerance beyond foliation. \href{https://doi.org/10.48550/arXiv.1810.09621}{arXiv:quant-ph/1810.09621} (2018).

\bibitem{LOCC}
E. Knill, R. Laflamme. A scheme for efficient quantum computation with linear optics. \href{https://doi.org/10.1038/35051009}{Nature 409, 7} (2001).

\bibitem{probabilistic_operation}
T. B. Pittman, B. C. Jacobs, J. D. Franson. Probabilistic Quantum Logic Operations Using Polarizing Beam Splitters. \href{https://doi.org/10.1103/PhysRevA.64.062311}{Phys. Rev. A 64, 062311} (2001).

\bibitem{loss_simulation}
M. Oszmaniec, D. J. Brod. Classical simulation of photonic linear optics with lost particles. \href{https://doi.org/10.1088/1367-2630/aadfa8}{New J. Phys. 20, 092002} (2018).

\bibitem{loss_simulation_nonuniform}
D. J. Brod, M. Oszmaniec. Classical simulation of linear optics subject to nonuniform losses. \href{https://doi.org/10.22331/q-2020-05-14-267}{Quantum 4, 267} (2020).

\bibitem{loss_effect}
J. E. Davis, D. Ö. Güney. Effect of loss on linear optical quantum logic gates. \href{https://doi.org/10.1364/JOSAB.430603}{J. Opt. Soc. Am. B, JOSAB 38, 153–159} (2021).

\bibitem{stace_loss_correction}
T. M. Stace, S. D. Barrett. Error Correction and Degeneracy in Surface Codes Suffering Loss. \href{https://doi.org/10.1103/PhysRevA.81.022317}{Phys. Rev. A 81, 022317} (2010).

\bibitem{stace_high_threshold}
S. D. Barrett, T. M. Stace. Fault tolerant quantum computation with very high threshold for loss errors. \href{https://doi.org/10.1103/PhysRevLett.105.200502}{Phys. Rev. Lett. 105, 200502} (2010).

\bibitem{herr_renormalization}
D. Herr, A. Paler, S. J. Devitt, F. Nori. A local and scalable lattice renormalization method for ballistic quantum computation. \href{https://doi.org/10.1038/s41534-018-0076-0}{npj Quantum Information 4, 1–8} (2018).

\bibitem{pant_thresholds}
M. Pant, D. Towsley, D. Englund, S. Guha. Percolation thresholds for photonic quantum computing. \href{https://doi.org/10.1038/s41467-019-08948-x}{Nature Communications 10, 1070} (2019).

\bibitem{kitaev_toric}
A. Yu. Kitaev. Quantum Error Correction with Imperfect Gates. \href{https://doi.org/10.1007/978-1-4615-5923-8\_19}{in: Hirota, O., Holevo, A.S., Caves, C.M. (Eds.), Quantum Communication, Computing, and Measurement. Springer US, Boston, MA, pp. 181–188.} (1997).

\bibitem{kitaev_codes}
A. Y. Kitaev. Quantum computations: algorithms and error correction. \href{https://doi.org/10.1070/RM1997v052n06ABEH002155}{Russ. Math. Surv. 52, 1191–1249} (1997).

\bibitem{planar_code}
S. Bravyi, A. Kitaev. Quantum codes on a lattice with boundary. \href{https://doi.org/10.48550/arXiv.quant-ph/9811052}{arXiv:quant-ph/9811052} (1998).

\bibitem{topological_quantum_memory}
E. Dennis, A. Kitaev, A. Landahl, J. Preskill. Topological quantum memory. \href{https://doi.org/10.1063/1.1499754}{Journal of Mathematical Physics 43, 4452–4505} (2002).

\bibitem{fowler_surface_codes}
A. G. Fowler, M. Mariantoni, J. M. Martinis, A. N. Cleland. Surface codes: Towards practical large-scale quantum computation. \href{https://doi.org/10.1103/PhysRevA.86.032324}{Phys. Rev. A 86, 032324} (2012).

\bibitem{floquet_QC}
S. Paesani, B. J. Brown. High-threshold quantum computing by fusing one-dimensional cluster states. \href{https://doi.org/10.1103/PhysRevLett.131.120603}{Phys. Rev. Lett. 131, 120603} (2022).

\bibitem{floquet_codes}
M. Davydova, N. Tantivasadakarn, S. Balasubramanian. Floquet codes without parent subsystem codes. \href{https://doi.org/10.1103/PRXQuantum.4.020341}{PRX Quantum 4, 020341} (2023).

\bibitem{color_code}
M. S. Kesselring, J. C. M. de la Fuente, F. Thomsen, J. Eisert, S. D. Bartlett, B. J. Brown. Anyon condensation and the color code. \href{https://doi.org/10.1103/PRXQuantum.5.010342}{PRX Quantum 5, 010342} (2024).

\bibitem{floquet_measurement}
J. R. Wootton. Measurements of Floquet code plaquette stabilizers. \href{https://doi.org/10.48550/arXiv.2210.13154}{arXiv:quant-ph/2210.13154} (2022).

\bibitem{zx_calculus_1}
B. Coecke, R. Duncan. Interacting Quantum Observables. \href{https://doi.org/10.1007/978-3-540-70583-3\_25}{in: Aceto, L., Damgård, I., Goldberg, L.A., Halldórsson, M.M., Ingólfsdóttir, A., Walukiewicz, I. (Eds.), Automata, Languages and Programming, Lecture Notes in Computer Science. Springer, Berlin, Heidelberg, pp. 298–310} (2008).

\bibitem{zx_calculus_2}
B. Coecke, R. Duncan. Interacting quantum observables: categorical algebra and diagrammatics. \href{https://doi.org/10.1088/1367-2630/13/4/043016}{New J. Phys. 13, 043016} (2011).

\bibitem{zx_calculus_review}
J. van de Wetering. ZX-calculus for the working quantum computer scientist. \href{https://doi.org/10.48550/arXiv.2012.13966}{arXiv:quant-ph/2012.13966} (2020).


\bibitem{teleportation}
C. H. Bennett, G. Brassard, C. Crépeau, R. Jozsa, A. Peres, W. K. Wootters. Teleporting an unknown quantum state via dual classical and Einstein-Podolsky-Rosen channels. \href{https://doi.org/10.1103/PhysRevLett.70.1895}{Phys. Rev. Lett. 70, 1895–1899} (1993).

\bibitem{quantum_gate_teleportation}
D. Gottesman, I. Chuang. Demonstrating the viability of universal quantum computation using teleportation and single-qubit operations. \href{https://doi.org/10.1038/46503}{Nature 402, 390–393} (1999).

\bibitem{qgt_universal}
M. A. Nielsen. Universal quantum computation using only projective measurement, quantum memory, and preparation of the 0 state. \href{https://doi.org/10.1016/S0375-9601(02)01803-0}{Physics Letters A 308, 96–100} (2003).

\bibitem{qgt_computation}
D. W. Leung. Quantum computation by measurements. \href{https://doi.org/10.48550/arXiv.quant-ph/0310189}{arXiv:quant-ph/0310189} (2004).

\bibitem{qgt_for_qc}
D. W. Leung. Two-qubit Projective Measurements are Universal for Quantum Computation. \href{https://doi.org/10.48550/arXiv.quant-ph/0111122}{arXiv:quant-ph/0111122} (2002).

\bibitem{qgt_fault-tolerant}
M. A. Nielsen, C. M. Dawson. Fault-tolerant quantum computation with cluster states. \href{https://doi.org/10.1103/PhysRevA.71.042323}{Phys. Rev. A 71, 042323} (2005).

\bibitem{qgt_derivation_for_mbqc}
A. M. Childs, D. W. Leung, M. A. Nielsen. Unified derivations of measurement-based schemes for quantum computation. \href{https://doi.org/10.1103/PhysRevA.71.032318}{Phys. Rev. A 71, 032318} (2005).

\bibitem{qgt_mbqc_intro}
R. Jozsa. An introduction to measurement based quantum computation. \href{https://doi.org/10.48550/arXiv.quant-ph/0508124}{arXiv:quant-ph/0508124} (2005).

\bibitem{aharonov_threshold}
D. Aharonov, M. Ben-Or. Fault-Tolerant Quantum Computation with Constant Error Rate. \href{https://doi.org/10.1137/S0097539799359385}{SIAM J. Comput. 38, 1207–1282} (2008).

\bibitem{fowler_threshold}
A. G. Fowler, A. M. Stephens, P. Groszkowski. High-threshold universal quantum computation on the surface code. \href{https://doi.org/10.1103/PhysRevA.80.052312}{Phys. Rev. A 80, 052312} (2009).

\bibitem{campbell}
E. T. Campbell, B. M. Terhal, C. Vuillot. Roads towards fault-tolerant universal quantum computation. \href{https://doi.org/10.1038/nature23460}{Nature 549, 172–179} (2017).

\bibitem{boosted_fusion}
W.P. Grice. Arbitrarily complete Bell-state measurement using only linear optical elements. \href{https://doi.org/10.1103/PhysRevA.84.042331}{Phys. Rev. A 84, 042331} (2011)


\bibitem{edmonds_1}
J. Edmonds. Maximum matching and a polyhedron with 0,
1-vertices. \href{https://doi.org/10.6028/JRES.069B.013}{Journal of research of the National Bureau of Standards
B, 69(125-130):55–56} (1965).

\bibitem{edmonds_2}
J. Edmonds. Paths, trees and flowers. \href{https://doi.org/10.4153/CJM-1965-045-4}{Canad. J. Math., 17:449} (1965).

\bibitem{pymatching}
O. Higgott, C. Gidney. Sparse Blossom: correcting a million errors per core second with minimum-weight matching. \href{https://doi.org/10.22331/q-2025-01-20-1600}{Quantum 9, 1600} (2025).

\bibitem{KLM}
E. Knill, R. Laflamme, and G. J. Milburn. A scheme for efficient quantum computation with linear optics. \href{https://doi.org/10.1038/35051009}{Nature 409, 46-52} (2001)

\bibitem{knill}
E. Knill. Quantum gates using linear optics and postselection. \href{https://doi.org/10.1103/PhysRevA.66.052306}{Phys. Rev. A 66, 052306} (2002)

\bibitem{pan-rudolph}
Q. Zhang, X.-H. Bao, C.-Y. Lu, X.-Q. Zhou, T. Yang, T. Rudolph, and J.-W. Pan. Demonstration of a scheme for the generation of “event-ready” entangled photon pairs from a single-photon source. \href{https://doi.org/10.1103/PhysRevA.77.062316}{Phys. Rev. A 77, 062316} (2008)

\bibitem{fldzhyan-bell}
S. A. Fldzhyan, M. Yu. Saygin, and S. P. Kulik. Compact linear optical scheme for Bell state generation. \href{https://doi.org/10.1103/PhysRevResearch.3.043031}{Phys. Rev. Research 3, 043031} (2021)

\bibitem{uskov2015}
D. B. Uskov, P. M. Alsing, M. L. Fanto, L. Kaplan, R. Kim, A. Szep, and A. M. Smith. Resource-efficient generation of linear cluster states by linear optics with postselection. \href{https://doi.org/10.1088/0953-4075/48/4/045502}{J. Phys. B: At. Mol. Opt. Phys. 48 045502} (2015)

\bibitem{wehner2019}
V. C. Vivoli, J. Ribeiro, and S. Wehner. High-fidelity Greenberger-Horne-Zeilinger state generation within nearby nodes. \href{https://doi.org/10.1103/PhysRevA.100.032310}{Phys. Rev. A 100, 032310} (2019)

\bibitem{varnava2008}
M. Varnava, D. E. Browne, and T. Rudolph. How Good Must Single Photon Sources and Detectors Be for Efficient Linear Optical Quantum Computation? \href{https://doi.org/10.1103/PhysRevLett.100.060502}{Phys. Rev. Lett. 100, 060502} (2008)

\bibitem{uskov2009}
D. B. Uskov, L. Kaplan, A. M. Smith, S. D. Huver, and J. P. Dowling. Maximal success probabilities of linear-optical quantum gates. \href{https://doi.org/10.1103/PhysRevA.79.042326}{Phys. Rev. A 79, 042326} (2009)

\bibitem{eisert2004}
J. Eisert. Optimizing Linear Optics Quantum Gates
. \href{https://doi.org/10.1103/PhysRevLett.95.040502}{Phys. Rev. Lett. 95, 040502} (2004)

\bibitem{gubarev}
F. V. Gubarev, I. V. Dyakonov, M. Yu. Saygin, G. I. Struchalin, S. S. Straupe, and S. P. Kulik. Improved heralded schemes to generate entangled states from single photons. 
\href{https://doi.org/10.1103/PhysRevA.102.012604}{Phys. Rev. A 102, 012604} (2020)

\end{thebibliography}

\appendix

\section{Simulation details}
\label{sec: simulation details}

Here, we present the error model and details of simulation that was used to estimate threshold parameters of the obtained four-qibit and ten-qubit FTFNs.
For any given Pauli observable measured within a fusion gate

\begin{itemize}
    \item the outcome of the observable measurement is lost with probability $p_{erasure}$,
    \item the outcome was not erased with probability $1-p_{erasure}$ and
    \begin{itemize}
        \item the outcome is flipped with probability $p_{error}$,
        \item it remains undisturbed with probability $1-p_{error}$.
    \end{itemize}
\end{itemize}

In the absence of errors, all fusion outcomes satisfy the parity condition expressed in the structure of the syndrome graph. In other words, the product of all outcomes of Pauli observables (which are either $+1$ or $-1$) associated with the edges that have the same origin must be equal to $1$. This means that an encoded state lays in the intersection of $+1$ eigenspaces of all elements of $\mathcal{C}=\mathcal{R}\cap \mathcal{F}$. However, a flip of a fusion outcome may cause a violation of the parity condition. In such case, some of the code stabilizers which are the generators of $\mathcal{C}=\mathcal{R}\cap \mathcal{F}$ produce an outcome $-1$. This set of check operators is called an error syndrome.

The problem of decoding implies the reconstruction of the error configuration that causes the observed error syndrome. We consider the decoding problem of topological codes represented in the syndrome graph formalism. Any error configuration can be represented as a set of chains, where each chain is a path consisting of edges corresponding to flipped outcomes. The ends of such a path indicate the error syndrome. The fusion outcomes pinpoint the vertices that correspond to the error syndrome. Our goal is to find the error configuration that causes this syndrome. We note that it is sufficient to restore the error chains up to a homological class \cite{topological_quantum_memory}.

Each edge of the syndrome graph is weighted proportionally to the flip probability of the corresponding fusion outcome. We restore the error chains using a decoder which employs a MWPM (Minimal Weight Perfect Matching) algorithm \cite{edmonds_1,edmonds_2} in order to connect all generated error syndromes with paths having a minimal total weight. If the composition of the restored error chain and the actual error chain forms a non-trivial cycle, then the decoding procedure is failed.

An erasure of the fusion outcomes leads to a removal of the corresponding edges in the syndrome graph. The vertices at the ends of the deleted edges should be merged. Thus, the erasure process modifies the syndrome graph, which is further analyzed by the MWPM decoder. It is important to note that if the erased edges percolate (i.e. in our case, form a continuous cluster of erased edges that includes a homologicaly nontrivial path), the decoding procedure cannot be successfully performed. In the implementation of the erasure simulation, there is no need to modify the actual graph's adjacency structure. Instead, we follow the approach presented in \cite{FBQC_1,stace_loss_correction}, where the weights of the erased edges are set to zero.

If the error probability is the same for all edge observables, one can set all weights of the edges of the syndrome graph to be equal to one. However, the syndrome graphs of the considered four-qubit and ten-qubit models contain multi-edges. In our simulations, we replace every multi-edge with a single edge. Every multi-edge can be considered as a composition of the corresponding Pauli observables and $p_{erasure}$ and $p_{error}$ must be updated. Indeed, if any element of a multi-edge is erased, then the entire composite observable is also erased. Then the probability of erasure transforms to

\begin{equation}
    p_{erasure}(m) = 1 - (1-p_{erasure})^m,
\end{equation}
where $m$ is the multiplicity of an edge. Finally, if an odd number of edges in multi-edge is flipped, then the associated composite observable is also flipped. Hence, for an edge of multiplicity $m$, we have

\begin{equation}
    p_{error}(m) = \frac{1-(1-2p_{error})^m}{2}.
\end{equation}

The dependence of these probabilities on parameter $m$ should be reflected in the choice of the edge weights. Luckily the MWPM decoder design easily handles such cases \cite{stace_loss_correction}. The weight of the edge is $\ln\left(\frac{1-p}{p}\right)$ if the error probability is $p$. Thus we get the weight of the edge with multiplicity $m$ to equal

\begin{equation}
    w(m) = \ln\left(\frac{1-p_{error}(m)}{p_{error}(m)}\right).
\end{equation}

Our goal is to estimate the fault tolerant region in the space of parameters $p_{error}$ and $p_{erasure}$. For points $(p_{error},\ p_{erasure})$ that lie inside this region, the probability of failure of the error correction procedure decreases exponentially with the distance from the boundary of the region. To estimate a point from the border of the fault tolerant region, we follow the strategy from the works \cite{FBQC_1,floquet_QC}. First, we choose a set of values $p_{error}$ and $p_{erasure}$ that lie in a line that comes through point $p_{error}=0,\ p_{erasure}=0$, i.e.

\begin{equation}
    p_{error}=c_{error}x,\quad p_{erasure}=c_{erasure}x.
\end{equation}

We repeat Monte-Carlo simulations of error and erasure noise for a constant pair of coefficients $(c_{error},\ c_{erasure})$ and several values of $x$ and perform the decoding procedure afterwards. We use $10^4$ samples to estimate the probability of successful correction $\Pi(x\ |\ c_{error},\ c_{erasure})$. Each sample is a simulated configuration of erased and flipped edges in the primal syndrome graph. Then, one calculates the error syndrome and performs the decoding procedure. The rate of successful decoding cases is regarded as an estimate of $\Pi(x\ |\ c_{error},\ c_{erasure})$. We use an MWPM-decoder based on sparse blossom algorithm implemented in Pymatching 2 library \cite{pymatching}.

We consider the syndrome graphs of the four-qubit and ten-qubit models as cubic lattices with periodic boundary conditions. Each lattice associates with a size parameter $L$, where $L^3$ is a number of vertices evenly spaced in a cube of size $L\times L\times L$. If we choose one corner of the cube as the coordinate origin, the locations of vertices can be presented in the form $(x,\ y,\ z)$, where $x,\ y,\ z=0,...,\ L-1$. In the case of the ten-qubit model, we connect graph nodes by edges using the following algorithm:

\begin{itemize}
    \item Each vertex $(x,\ y,\ z)$ is connected with $4$ vertices $(x\pm 1,\ y,\ z)$ and $(x,\ y\pm 1,\ z)$. These edges have the weight equal to $w(1)$.
    \item If $x+y$ is an even number, then the vertex is additionally connected with $2$ vertices $(x,\ y,\ z\pm1)$ by the weight $w(4)$ edges.
\end{itemize}
To provide periodic boundary conditions, the calculations are performed modulo $L$. For the four-qubit model we have

\begin{itemize}
    \item Each vertex $(x,\ y,\ z)$ is connected with $4$ vertices $(x\pm 1,\ y,\ z)$ and $(x,\ y\pm 1,\ z)$. These edges has the weight equal to $w(1)$.
    \item If $x+y$ is an even number, then the vertex is additionally connected with $2$ vertices $(x,\ y,\ z\pm1)$ by edges the weight $w(4)$ edges, with $4$ vertices $(x\pm 1,\ y\pm 1,\ z)$ by the weight $w(3)$ edges and by the weight $w(1)$ edges with $4$ vertices located in
    \begin{itemize}
        \item $(x\pm 1,\ y\pm 1,\ z+1)$ if $y$ is even,
        \item $(x\pm 1,\ y\pm 1,\ z-1)$ if $y$ is odd.
    \end{itemize}
\end{itemize}
Here, we also perform all calculations modulo $L$. Note, according to the present instructions of the syndrome graphs construction value $L$ must be even.

For each set of parameter values $c_{error},\ c_{erasure},\ x$ we perform simulations for $L=12,\ 16,\ 20$. Thus, for each $c_{error}$ and $c_{erausre}$, we obtain three curves $\Pi_L(x\ |\ c_{error},\ c_{erasure})$. We search a threshold value of $x=x_{th}$ as an intersection of these curves. Finally, we obtain a point $(c_{error}x_{th},\ c_{erasure}x_{th})$ from the border of the fault tolerant region.

\begin{figure}
\centering
    \includegraphics[scale=0.6]{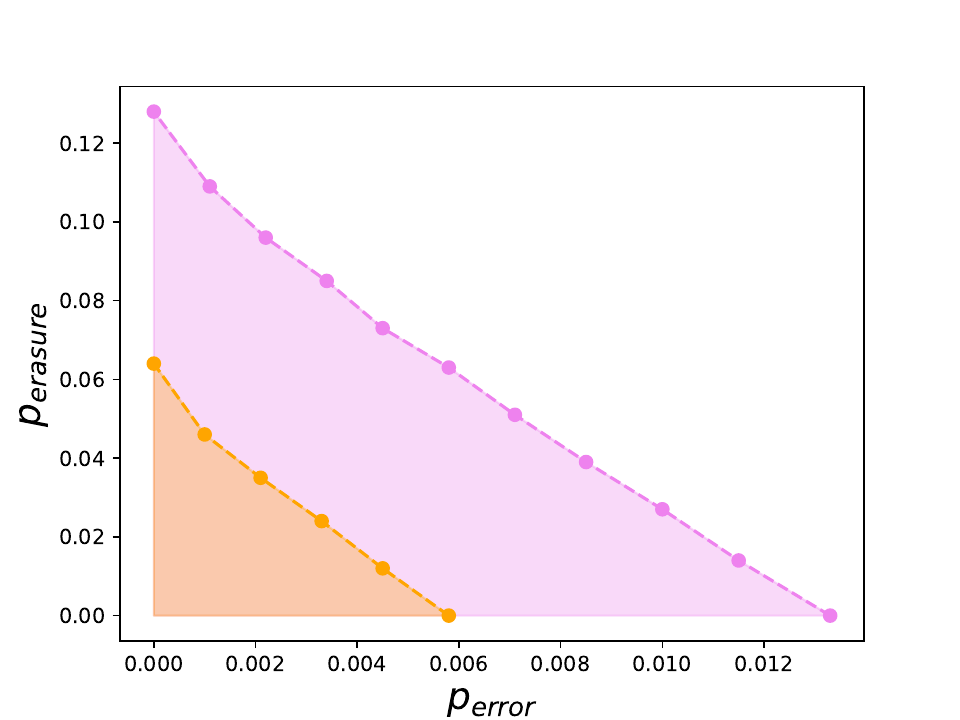}
    \caption{Regions of correctable noise rates for the 4-qubit model (orange) and for the 10-qubit model (violet).}
    \label{fig:thresholds}
\end{figure}

By varying the values of the parameters $c_{error}$ and $c_{erasure}$, we estimate a curve that defines the boundary of the fault-tolerant region in the space of $(p_{error},\ p_{erasure})$, see Fig.~\ref{fig:thresholds}.

\end{document}